\documentclass[twocolumn,aps,prd]{revtex4-1}
\usepackage{slashed}
\usepackage{enumitem}

\usepackage{float}

\usepackage{tikz-cd}
	\usepackage{amsmath}
	\usepackage{amsfonts}
	\usepackage{amssymb}
	\usepackage{graphicx}
	\usepackage{mathtools}
	\usepackage{stmaryrd}
	\usepackage{autobreak} %for long equation
    \usepackage{cancel}
 
	\allowdisplaybreaks
	\usepackage[colorlinks=true,citecolor=blue,linkcolor=red]{hyperref}
	\usepackage{bbold}					% for \mathbb{1} as a nice unit matrix symbol
	\usepackage{multirow}				% merge cells vertically in tables
	\usepackage[normalem]{ulem}        % strike-through text
	\usepackage{array} 
    \usepackage{cancel} 
\usepackage{lipsum}
\usepackage{graphicx}

\usepackage{titletoc} %creat table of content  for appendix
\usepackage{mathtools}

	\newcolumntype{x}[1]{>{\centering\let\newline\\\arraybackslash\hspace{0pt}}p{#1}}
	
	\makeatletter
\newcommand*\rel@kern[1]{\kern#1\dimexpr\macc@kerna}
\newcommand*\widebar[1]{%
  \begingroup
  \def\mathaccent##1##2{%
    \rel@kern{0.8}%
    \overline{\rel@kern{-0.8}\macc@nucleus\rel@kern{0.2}}%
    \rel@kern{-0.2}%
  }%
  \macc@depth\@ne
  \let\math@bgroup\@empty \let\math@egroup\macc@set@skewchar
  \mathsurround\z@ \frozen@everymath{\mathgroup\macc@group\relax}%
  \macc@set@skewchar\relax
  \let\mathaccentV\macc@nested@a
  \macc@nested@a\relax111{#1}%
  \endgroup
}
\makeatother

	\DeclareMathOperator{\sign}{sign}  	% function \sign{x}
	\DeclareMathAlphabet{\mathbbold}{U}{bbold}{m}{n}
	\newcounter{subeqn} %
	\makeatletter
	\@addtoreset{subeqn}{equation}
	\makeatother

\definecolor{ZM}{rgb}{.5,0,.5}
\definecolor{SD}{rgb}{0,1,0}

\newcommand\trick[1]{} %For equations in the footnote end with a period. add "\protect\trick." at the end of footnote.

\begin{document}
\title{Coherent information as a mixed-state topological order parameter of fermions}

\author{Ze-Min Huang$^1$}
\author{Luis Colmenarez$^{2, 3}$}
\author{Markus M\"{u}ller$^{2, 3}$}
\author{Sebastian Diehl$^1$}

\affiliation{$^1$Institute for Theoretical Physics, University of Cologne, 50937 Cologne, Germany}
\affiliation{$^2$Institute for Quantum Information, RWTH Aachen University, 52056 Aachen, Germany}
\affiliation{$^3$Institute for Theoretical Nanoelectronics (PGI-2), Forschungszentrum Jülich, 52428 Jülich, Germany}

\begin{abstract}
Quantum error correction protects quantum information against decoherence provided the noise strength remains below a critical threshold. This threshold marks the critical point for the decoding phase transition. Here we connect this transition in the toric code to a topological phase transition in disordered Majorana fermions at high temperatures. A quantum memory in the error correctable phase is captured by the presence of a Majorana zero mode, trapped in vortex defects associated with twisted boundary conditions. These results are established by expressing the coherent information, which measures the amount of recoverable quantum information in a given noisy code, in terms of a mixed-state topological order parameter of fermions. Our work hints at a broader connection of the robustness of quantum information in stabilizer codes and mixed-state topological phase transitions in symmetry protected fermion matter. 
\end{abstract}

\maketitle
%\tableofcontents

\section{Introduction }

Quantum states are susceptible to decoherence and noise, which can introduce errors and drive the system toward a mixed state, thereby degrading the quantum information content. Quantum error correction (QEC) is thus essential to preserve quantum information and protect it against errors~\cite{nielsen2010cambridge, gottesman2010intro,terhal2015rmp}. This process involves encoding logical qubits into entangled states across multiple physical qubits, making logical states thereby indistinguishable at the local level. However, this protection of logical quantum information against noise lasts only up to a certain code- and noise-model dependent error threshold, beyond which any decoding algorithm will fail, thus marking a decoding phase transition \cite{dennis2002jmp, wang2003aop, bombin2008pra, katzgraber2009prl,bombin2012prx, kubica2018prl,nigg2014science, vodola2022quantum, song2022prl}. This transition is pivotal for determining the boundary of functioning fault-tolerant quantum computing \cite{nielsen2010cambridge, gottesman2010intro,terhal2015rmp}, making its study critical for advancing practical scalable quantum information processing. 

A classic result in QEC concerns the critical error threshold of the toric code under bit or phase flip errors: here, the error threshold of the noisy QEC code has been connected to the phase transition in the random bond Ising model (RBIM)~\cite{dennis2002jmp}.~ Subsequent works have established similar connections to classical disordered statistical mechanics models for other QEC codes and noise models \cite{ wang2003aop, bombin2008pra, katzgraber2009prl,bombin2012prx, kubica2018prl, vodola2022quantum, song2022prl}. Recently, the connection between quantum information and statistical mechanics models has been further strengthened: Reference~\cite{fan2024prxQ} demonstrates a direct mapping of information theoretic quantities to the RBIM. One of these is the coherent information (CI), a measure for the amount of recoverable information within logical qubits under decoherence \cite{schumacher1996pra, lyoyd1997pra}.
A classic result in condensed matter physics, on the other hand, concerns the mapping of the RBIM to disordered Majorana fermions \cite{dotsenko1983ap, cho1997prb, shalaev1994pr, read2000prb, gruzberg2001prb, merz2002prb,wille2024prr} in symmetry class D \cite{altland1997prb}, the one with the least amount of physical symmetries: number conservation, time reversal, and charge conjugation are all absent. This mapping has been established only for the bulk of the system, and in the thermodynamic limit. Nevertheless, this mapping suggests a connection of the information content in the toric code to the physics of topological superconductors. 

\begin{figure}[th!]
\includegraphics[scale=0.20]{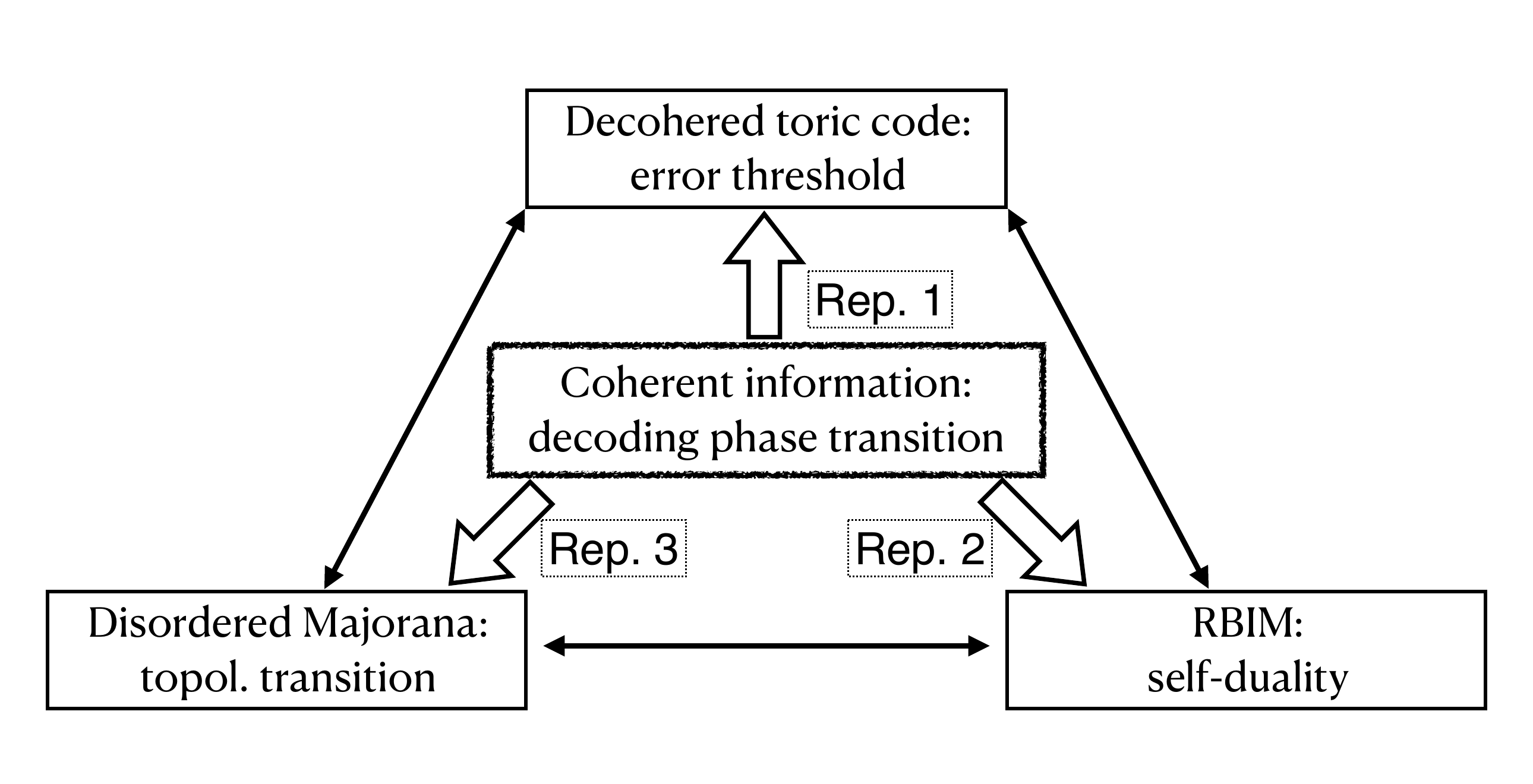}

\caption{Overview of different representations of the coherent information (CI) for the toric code under both bit-flip and phase errors. Here, the CI measures the residual information that is recoverable under decoherence, with the zero point of the CI marking the decoding phase transition. This point acquires different physical interpretations in different representations: It corresponds to the critical error threshold in the decohered toric code (representation 1), and the low/high temperature self-duality point in the random-bond Ising model (RBIM, representation 2)~\cite{fan2024prxQ}. In this work, we create a link to symmetry-protected topological quantum matter in terms of disordered Majorana fermions (representation 3). In this representation, the coherent information is directly tied to a mixed-state topological order parameter of fermions under twisted boundary conditions. The decoding phase transition is described by a topological phase transition of fermions in a mixed quantum state. A non-trivial quantum memory in the error correctable phase is captured by the presence of a Majorana zero mode, induced by vortex defects associated to the boundary conditions. For a further summary of results, see Sec.~\ref{sec:TSC} and Fig.~\ref{fig:braiding}.
\label{fig:concepts}}
\end{figure}

In this work, we uncover this connection and make it precise, based on a mapping of the CI to disordered Majorana fermions, which is exact for any system size $N$ (i.e., the number of physical qubits in the error correction code). More broadly, this connects the various above threads under the common umbrella of the coherent information (CI), giving rise to the triptych displayed in Fig.~\ref{fig:concepts}. In particular, the decoding phase transition in the toric code corresponds to a topological phase transition in the fermion system in the following way:
\begin{enumerate}
    \item The error correctable phase coincides with the Majorana topological phase. 
   \item The code space with recoverable quantum information content corresponds to the presence of a Majorana zero mode due to bulk-vortex correspondence.
   \item The error threshold equals the point of zero vortex fugacity.
\end{enumerate}
The key technical ingredient enabling these statements is the exact nature of the mapping of the CI for any finite $N$, but also in the thermodynamic limit $N\to\infty$. Importantly, this allows us to carefully track the boundary conditions of the decohered toric code, which encode the quantum information content measured by the CI. In the Majorana representation, these boundary conditions surface in a way which allows us to interpret the coherent information as a mixed-state topological order parameter of fermions \cite{huang2024arxiv}. Specifically, this order parameter is given by the expectation value of the fermion parity operator of a finite-temperature, disordered Majorana system under twisted spatial boundary conditions. This parity probe operator can be viewed as a temporal boundary condition in a finite temperature partition function, corresponding to the insertion of a temporal $\mathbb{Z}_2$ gauge flux. The spatial boundary conditions instead correspond to the insertion of a spatial $\mathbb{Z}_2$ gauge flux. They create topological defects such as vortices, in turn trapping Majorana zero modes in the topologically non-trivial, respectively error correctable phase.

These findings create a new link between the robustness of quantum information and the mixed-state topology of fermions. Three features of the phase transition in the fermion representation are particularly worth pointing out: (i) The mixed-state topological transition proceeds without any thermodynamic singularities, such as the divergence of a correlation length. Rather, its defining feature is the loss of two Majorana zero modes, once the critical error threshold of the quantum memory is exceeded. This rationalizes the loss of two qubits of information as a boundary effect — the two logical qubits of the toric code are lost in a bulk of $N$ physical qubits — while the extensive thermodynamics remains unaffected thereby. (ii) The thermodynamic limit $N\to \infty$ is needed to suppress hybridization of Majorana zero modes trapped in defects. This corresponds to the sealing of quantum information to two logical qubits, which occurs only in the thermodynamic limit.  
(iii)  The mixed-state topological order parameter — the expectation value of the fermion parity operator — is linear in the state, but global in nature. In this way, it is able to resolve the non-locally encoded information. In this representation, it is thus not the non-linearity in the density matrix that matters for the resolution of the decoding transition (as is the case for the CI in terms of the original toric code density matrix), but rather the ability to formulate order parameters that capture the topological properties of the state.

\begin{figure}
\centering
\includegraphics[scale=0.18]{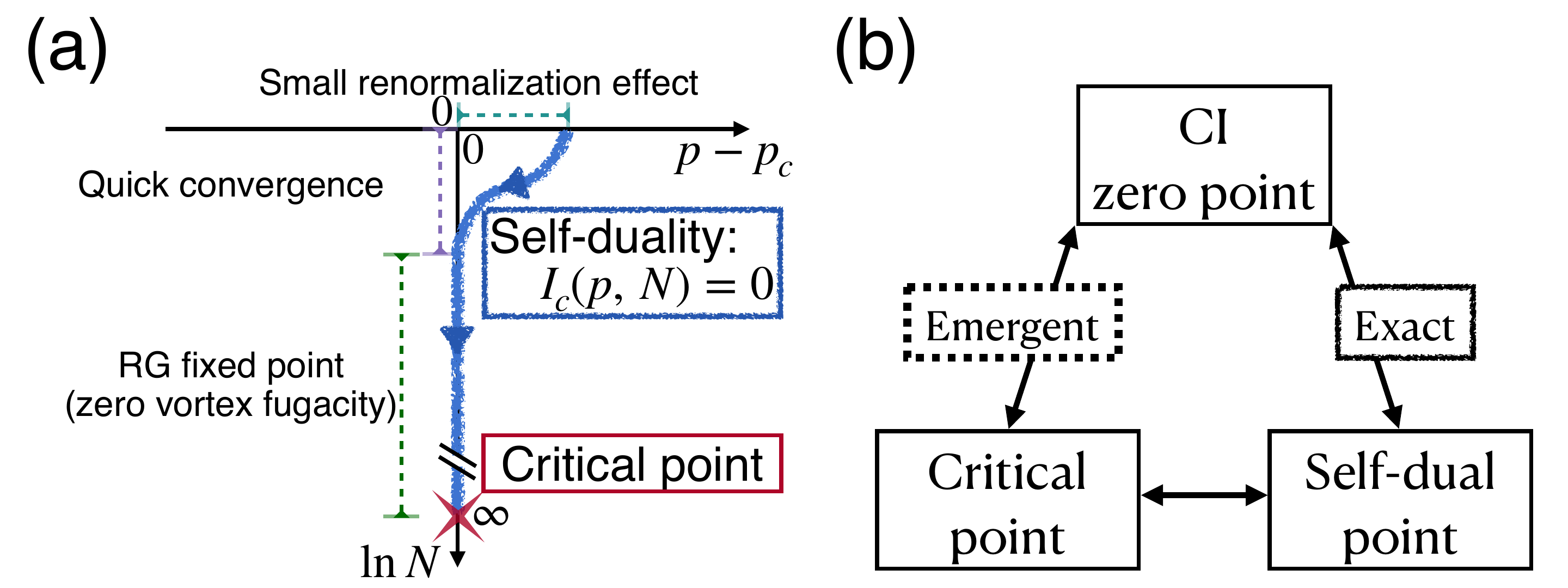}
\caption{Summary of the relation between the 
coherent information (CI) zero crossing point, the critical point, and the self-dual point. The CI zero point ($I_c(p,\ N)=0$) can depend on both the error rate $p$ and the system size $N$, as schematically illustrated by the blue line in panel (a). Mapping the decohered toric code to a disordered Majorana model reveals the physical meaning of $I_c=0$ as the zero vortex fugacity point. The zero vortex fugacity point rapidly converges to a renormalization group (RG) fixed point, leading to an \textit{emergent} coincidence between the zero CI point and the critical point in the thermodynamic limit ($p=p_c$, red star in (a)). Additionally, we find an \textit{exact} correspondence between the zero CI point and the self-duality point in the random-bond Ising model (RBIM) for any system size $N$ (blue frame in (a)), where the RBIM arises from a statistical mechanics mapping of the CI. Together, these results establish the relation between the zero CI point, the critical point, and the self-dual point, as shown in panel (b).
\label{fig:3points}}
\end{figure}

Furthermore, resulting from these findings, we obtain two important corollaries, see Fig.~\ref{fig:3points}. The first of these exports insights from many-body theory to quantum information: The tiny finite-size effects of the error threshold observed in the toric code are rationalized from the above highlighted  connection to the zero vortex fugacity point: As established in quantum Hall physics, this is a renormalization group fixed point, to which the flow converges rapidly, and becomes size-insensitive (Fig.~\ref{fig:3points} (a))~\cite{khmelnitskii1983jetp, pruisken1984npb, levine1984npb, huckestein1995rmp, read2000prb, fu2012prl, konig2012prb, altland2014prl, altland2015prb}. The connection to universal renormalization group flows suggests that the small finite-size effects should also be present in more general stabilizer codes, and we provide numerical evidence for this hypothesis \cite{fnCI}. In the context of practical QEC this corroborates that the CI constitutes a powerful tool to accurately determine fundamental critical error thresholds of noisy QEC codes already from small (i.e.~low distance) code instances, as observed in~\cite{colmenarez2024prr}.
The second corollary operates in reverse (Fig.~\ref{fig:3points} (b)): Our results on the CI imply that the critical point in the RBIM is self-dual. This follows from combining the two independently derived facts that (i) for arbitrary system size $N$, the zero crossing point of the CI  coincides with the self-dual point in the RBIM, and (ii), in thermodynamic limit, that zero crossing coincides with the critical point of this model (also that of the Majorana model, but relevant here is the former). While this connection has been conjectured for a long time \cite{nishimori2007jsp}, its demonstration, enabled by the connection to the information theoretic CI, is new to the best of our knowledge.
 
This paper is organized as follows: In Sec.~\ref{sec:qec_CI}, we review quantum error correction using the CI, and apply it to two examples: (1) a decohered qubit and (2) the decohered toric code, mapped to the RBIM. In Sec.~\ref{sec:dToric_CI}, we show that the CI zero point identifies the self-dual point of the RBIM. In Sec.~\ref{sec:TSC}, we introduce the
Majorana representation for the CI, and reveal
its connection to the mixed-state topological order parameter. In Sec.~\ref{sec:numerical_result}, we present numerical results for the CI in other topological stabilizer codes. We conclude in Sec.~\ref{sec:conclude_outlook}.

\section{Quantum error correction via coherent information \label{sec:qec_CI}}

A key concept for the present work is the coherent information (CI) \cite{schumacher1996pra, lyoyd1997pra, barnum1998pra, nielsen2010cambridge, colmenarez2024prr, zhao2023arxiv, lee2024arxiv, lyons2024arxiv, niwa2024arxiv, hauser2024arxiv, chen2024arxivCI, eckstein2024prxq}, denoted by
$I_{c}$. It quantifies the
information of a quantum code retained after exposure to noise. Therefore, $I_c$ determines error correctability \cite{schumacher1996pra, barnum1998pra}: Quantum error correction is achievable if and only if the residual information ($I_c$) equals the initially stored information ($S_Q$), a condition equivalent to the Knill-Laflamme criterion (see Appendix \ref{supp_sec:EC_condition_CI} for details). Importantly, $I_c$ decreases monotonically with the error rate, reflecting increased information leakage through noisy channels in line with the quantum data processing inequality  \cite{nielsen2010cambridge}.

We next provide the quantitative definition of CI, and apply it to a single qubit undergoing bit-flip and phase errors, for an illustration. We then extend this to the decohered toric code, reviewing its connection to the random-bond Ising model (RBIM) \cite{dennis2002jmp, fan2024prxQ}.

\subsection{Coherent information: Generalities}
\begin{figure}[t!]
\includegraphics[scale=0.18]{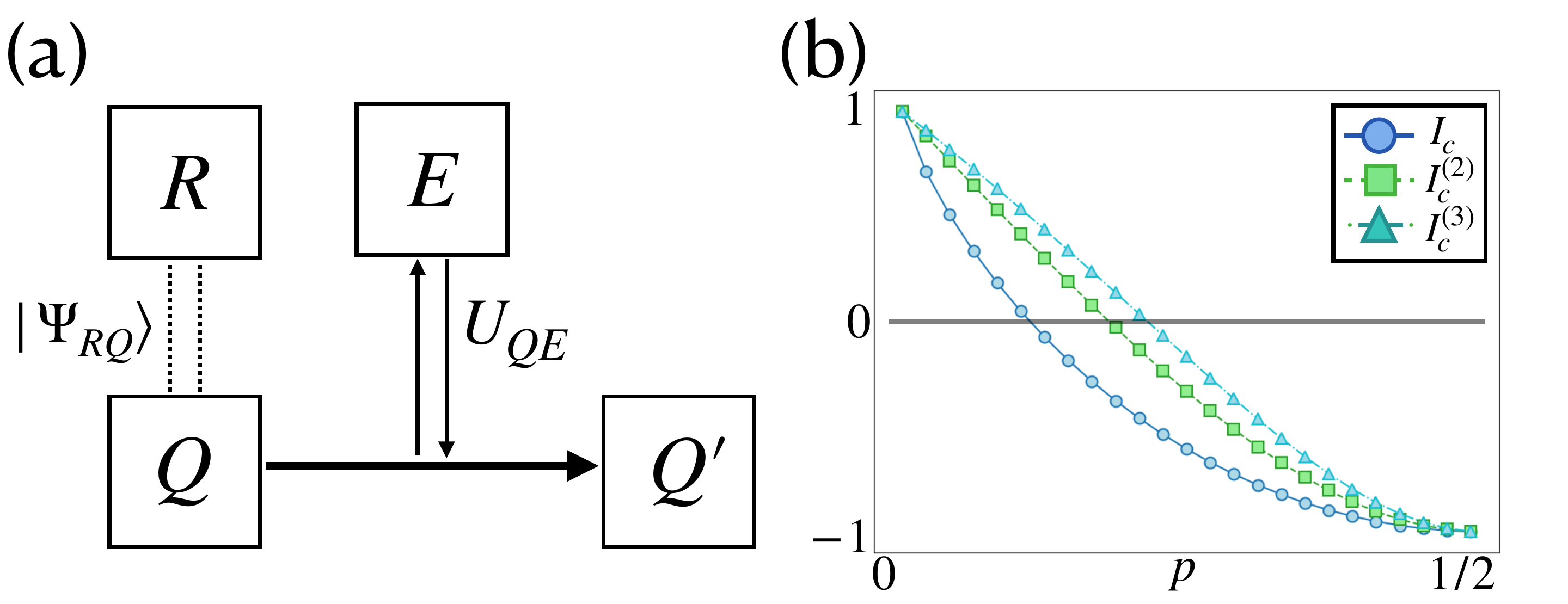}

\caption{Coherent information setup (a) and its dependence on the error
rate $p$ for a single qubit under bit-flip and phase errors (b).
In (a), $Q$ represents the quantum memory with density matrix $\rho_{Q}$, purified to $|\Psi_{RQ}\rangle$ by introducing a reference qubit
$R$, i.e., $\rho_{Q}=\text{Tr}_{R}\left(|\Psi_{RQ}\rangle\langle\Psi_{RQ}|\right)$.
Decoherence occurs due to coupling with the environment $E$, evolving
the initial state $|\Psi_{RQ}\rangle\otimes|0_{E}\rangle$ to $|\Psi_{RQ^{\prime}E^{\prime}}\rangle=U_{QE}|\Psi_{RQ}\rangle\otimes|0_{E}\rangle$. In (b), $I_c$ and $I_c^{(2,\ 3)}$ represent the coherent information and its R\'enyi-$2, 3$ counterparts, respectively, both of which decrease monotonically with $p$.
\label{fig:CI_1Qubit}}
\end{figure}

We define the CI for a quantum code $Q$ using the setup in Fig. \ref{fig:CI_1Qubit} (a), and derive the formula for the error correction condition. To assess the information in $Q$, we introduce a reference system $R$, which we maximally entangle with $Q$; we denote this state by $|\Psi_{RQ}\rangle$. We then define the reduced density matrices $\rho_{R/Q}$ of subsystems $R/Q$ and their von Neumann entropies before the interaction with the environment, 
\begin{eqnarray}
\rho_{Q/R}=\text{Tr}_{R/Q}\left(|\Psi_{RQ}\rangle\langle\Psi_{RQ}|\right).
\end{eqnarray}
The von Neumann entropy of $Q$ is
\begin{eqnarray}
S_{Q}\equiv-\text{Tr}_{Q}(\rho_Q \log_2 \rho_Q) =S_{R},
\end{eqnarray}
where $S_{R}\equiv-\text{Tr}_R\left(\rho_R\log_2 \rho_R\right)$. The equality $S_Q=S_R$ holds since $|\Psi_{RQ}\rangle$ is pure, so $S_R$ reflects the initially stored information.  An alternative way of looking at the situation is to say that $|\Psi_{RQ}\rangle$ is a purification of $\rho_Q$.

During the  interaction with the environment,  information may leak to the latter, captured by the quantum mutual information between $R$ and $E^\prime$,
\begin{equation}
I_m(R: E^\prime) \equiv S_R +S_{E^\prime} -S_{RE^\prime}.
\end{equation}
where primes (e.g., $E^\prime$) indicate post-interaction states.
Specifically, interaction evolves $Q$ and $E$ into $Q^\prime$ and $E^\prime$ via a unitary transformation $U_{QE}$,
\begin{equation}
\rho_{Q^{\prime}/E^\prime/RQ^\prime}\equiv\text{Tr}_{R,E/R, Q/ E^\prime}\left[U_{QE}\left(|\Psi_{RQE}\rangle\langle\Psi_{RQE}|\right)U_{QE}^{\dagger}\right],
\end{equation}
where the slash notation in the subscript (e.g., $Q^{\prime}/E^\prime/RQ^\prime$) indicates that the expression applies to any of the options separated by the slash. $|\Psi_{RQE}\rangle\equiv |\Psi_{RQ}\rangle\otimes |0_E\rangle$ as there is no initial entanglement between $RQ$ and $E$. $\rho_R$ remains unchanged since $U_{QE}$ acts only on $Q$ and $E$. 
The CI, measuring how much information remains in $Q$, is thus defined by subtracting the leaked information ($I_m$) from the initially stored one ($S_R$), 
\begin{equation}
I_c \equiv S_{R} - I_m(R:E^\prime) = S_{Q^\prime}-S_{RQ^\prime}, \label{eq:coherent_info}
\end{equation}
where the second equality follows from the purity of the state $U_{QE}|\Psi_{RQE}\rangle$, i.e., $S_{E^\prime} = S_{RQ^\prime}$ and $S_{RE^\prime} = S_{Q^\prime}$.
Thus, successful quantum error correction is possible if and only if,
\begin{equation}
I_{c}=S_{Q}, \label{eq:QEC_CI}
\end{equation}
indicating the stored information remains intact despite the noise.

We then utilize the replica trick to compute the von Neumann entropy $S_{Q^\prime}$ and $S_{RQ^\prime}$ of the CI, illustrated in the single qubit example below, 
\begin{equation}
I_c = \lim_{n\rightarrow 1^+} I_c^{(n)},
\end{equation}
where $n$ is the replica index. The R\'enyi-$n$ CI is defined as 
\begin{equation}
\begin{cases}
I_{c}^{(n)}=S_{Q^{\prime}}^{(n)}-S_{RQ^{\prime}}^{(n)},\\
S_{RQ^{\prime}/Q^{\prime}}^{(n)}= - \frac{1}{n-1}\log_{2}\text{Tr}\rho_{RQ^{\prime}/Q^{\prime}}^{n}.
\end{cases}
\end{equation}
We now evaluate the R\'enyi-$n$ CI $I_c^{(n)}$ and $I_c$ for a single qubit under a bit-flip and a phase channel with error rate $p$, elucidating two key tools used later on --  the replica trick and the monotonic behavior of the CI $I_c$ as a function of $p$. We start from a maximally entangled Bell state, 
$
|\Psi_{RQ}\rangle=\left(|0_{Q}0_{R}\rangle+|1_{Q}1_{R}\rangle\right)/\sqrt{2}$,
where $Q$ (system) and $R$ (reference) are both single qubits.
Working in the stabilizer formalism, we have
\begin{equation}
\rho_{RQ}=\frac{1}{4}\left(\mathbb{I}+\sigma_{Q}^{x}\otimes\sigma_{R}^{x}\right)\left(\mathbb{I}+\sigma_{Q}^{z}\otimes\sigma_{R}^{z}\right),\ \ \rho_{Q}=\frac{1}{2}\mathbb{I}.
\end{equation}
Applying the bit-flip and phase channels, we get
\begin{equation}
\rho_{RQ^{\prime}/Q^\prime}  =  \mathcal{E}_{Z}\circ\mathcal{E}_{X}\left[\rho_{RQ/Q}\right],
\end{equation}
where 
\begin{equation}\label{eq:noise0}
%\begin{cases}
    \mathcal{E}_{X/Z}[\rho]= (1-p)\ \rho + p\ \sigma^{x/z}_Q\ \rho\ \sigma^{x/z}_Q, % \\
%    \mathcal{E}_{Z}[\rho_{RQ/Q}]= (1-p)\ \rho_{RQ/Q} + p\ \sigma^{z}_Q\ \rho_{RQ/Q}\ \sigma^{z}_Q
%\end{cases}
\end{equation}
and $\sigma^{x/z}_Q$ are Pauli matrices acting on $Q$.  This yields
\begin{equation}
S^{(n)}_{RQ^{\prime}}=-\frac{2}{n-1}\log_{2}\left[\left(1-p\right)^{n}+p^{n}\right], \,\, S_{Q^{\prime}}^{(n)}=1,
\end{equation}
and thus we obtain
\begin{equation}
I_{c}^{\left(n\right)}  =  \frac{2}{n-1}\log_{2}\left[\left(1-p\right)^{n}+p^{n}\right]+1,\label{eq:CI_n_1qubit}
\end{equation}
In the $n\rightarrow 1^+$ limit, we find the CI
\begin{equation}
I_{c}=2\left[p\log_{2}p+\left(1-p\right)\log_{2}\left(1-p\right)\right]+1.\label{eq:CI_1qubit}
\end{equation}
Clearly, $I_{c}$ decreases monotonically with increasing error rate
(see Fig.~\ref{fig:CI_1Qubit} (b)), in accordance with the quantum data processing inequality \cite{nielsen2010cambridge}.

\subsection{Coherent information for the decohered toric code
\label{sec:dToric_RBIM}}
We next apply the CI concept to the toric code, a classic example
of topological order and quantum memory~\cite{fan2024prxQ}. The toric code is a many-body
system with degenerate ground states forming a robust code space consisting of two logical qubits,
where logical operations are executed by non-local operators, as local
perturbations cannot cause change between ground states. Calculating
the CI Eq. (\ref{eq:coherent_info}) requires therefore two reference
qubits, each maximally entangled with one of the non-local logical qubits. Here we will elaborate that this  process links to twisted boundary conditions (or equivalently, flux insertion) in a many-body system. This perspective will be leveraged to -- and fruitful for -- the fermion representation constructed in the subsequent section 

\subsubsection{Coherent information for the toric code \label{subsubsec:review_DeTC}}

\begin{figure}
\includegraphics[scale=0.18]{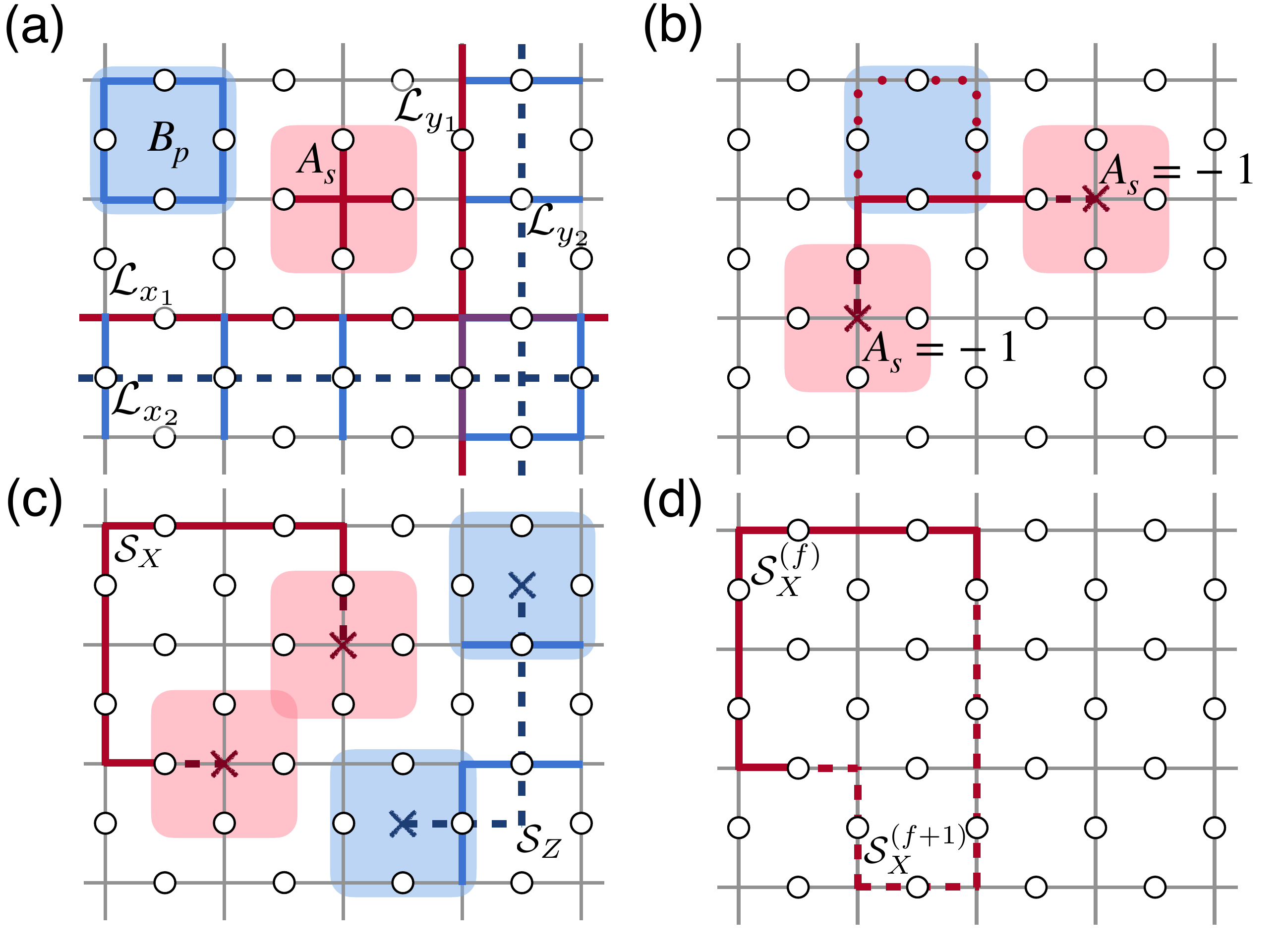}

\caption{Illustration of the toric code: (a) shows the stabilizers $A_{s}$ and
$B_{p}$ in the toric code Hamiltonian (Eq.~\eqref{eq:Ham_toric}), as well as the non-contractible loops in the $x$- and $y$- direction, $\mathcal{L}_{x_{1/2}}$ and $\mathcal{L}_{y_{1/2}}$. (b) depicts a string excitation
with $A_{s}=-1$ at both ends. Applying $B_{p}$ (blue shaded area),
changes the string's shape, while keeping the ends intact. (c) plots error
chains, or string excitations created by chains of bit-flip ($\mathcal{S}_{X}$)
or phase ($\mathcal{S}_{Z}$) errors. The error corrupted density matrix $\rho_{Q^\prime/RQ^\prime}$ (Eq.~\eqref{eq:rho_RQ_p}) represents a weighted ensemble of these chains. (d) illustrates that in the
R\'enyi-$n$ entropy (used for the R\'enyi-$n$ CI in Eq.~\eqref{eq:Renyi_n_CI_2qubit}), an error chain (e.g., $\mathcal{S}_{X}$), contributes
only when $\mathcal{S}_{X}^{\left(f\right)}$ and $\mathcal{S}_{X}^{\left(f+1\right)}$
in adjacent replicas form a closed loop. \label{fig:toric_code}}
\end{figure}

We briefly recapitulate some basics of the toric code, and define the corresponding density matrix with/without reference qubits, $\rho_{RQ/Q}$ (see Eq. \eqref{eq: rho_RQ_TC} and Eq. \eqref{eq: rho_Q_TC}). Its code space is spanned by four degenerate ground states of the Hamiltonian
on a square lattice with $N$ sites and periodic boundary conditions, which provides a quantum memory. The Hamiltonian is \cite{kitaev2003aop}
\begin{equation}\label{eq:Ham_toric}
H_{\text{TC}}=-\sum_{s}A_{s}-\sum_{p}B_{p},
\end{equation}
where the subscripts '$s$' and '$p$' denote vertices (or stars) and
plaquettes, respectively (see Fig. \ref{fig:toric_code} (a)). The
operators $A_{s}$ and $B_{p}$ commute and are defined as:
\begin{equation}
\begin{cases}
A_{s}=\prod_{l\in s}\sigma_{l}^{z}\\
B_{p}=\prod_{l\in p}\sigma_{l}^{x}
\end{cases},
\end{equation}
with the Pauli matrices $\sigma_{l}^{x/z}$ residing on the bond
 $l$. These operators satisfy
\begin{equation}
\prod_{\forall s}A_{s}=\prod_{\forall p}B_{p}=\mathbb{I},\label{eq:A_s_B_p}
\end{equation}
indicating that only $2N-2$ of $A_{s}$ and $B_{p}$ are independent.
Thus, the ground states of $H_{\text{TC}}$ span a four-dimensional
code space, as they are the $+1$ eigenvectors of $A_{s}$ and $B_{p}$,
leaving $4=\frac{2^{2N}}{2^{2N-2}}$ states unconstrained. The numerator
$2^{2N}$ represents the Hilbert space dimension, while the denominator
$2^{2N-2}$ accounts for the constraints from $A_{s}/B_{p}$. Thus, the code space encodes two logical qubits with logical operators, 
\begin{equation}
\begin{cases}
X_{1}\equiv\prod_{l\in\mathcal{L}_{x_1}}\sigma_{l}^{x},\ Z_{1}\equiv\prod_{l\in\mathcal{L}_{y_2}}\sigma_{l}^{z}\\
X_{2}\equiv\prod_{l\in\mathcal{L}_{y_1}}\sigma_{l}^{x},\ Z_{2}\equiv\prod_{l\in\mathcal{L}_{x_2}}\sigma_{l}^{z}
\end{cases},
\end{equation}
where $\mathcal{L}_{x_{1/2}/y_{1/2}}$ are non-contractible loops along the $x/y$
direction of the torus (see Fig. \ref{fig:toric_code} (a)), arising from the symmetries of $H_{\text{TC}}$, i.e., $\left[H,\ X_{1 / 2}\right]=\left[H,\ Z_{1 / 2}\right]=0$. 

The density matrix $\rho_{RQ}$ is defined by maximally entangling the logical qubits with reference qubits $R$, giving 
\begin{equation}\label{eq: rho_RQ_TC}
\begin{cases}
\rho_{RQ}=\rho_{R_{1}Q}^{0}\times\rho_{R_{2}Q}^{0}\times\left(\prod_{\forall s}P_{s}\right)\times\left(\prod_{\forall p}P_{p}\right)\\
\rho_{R_{1/2}Q}^0\equiv\frac{1}{4}(\mathbb{I}+X_{1/2}\otimes\sigma_{R_{1/2}}^{x})(\mathbb{I}+Z_{1/2}\otimes\sigma_{R_{1/2}}^{z})
\end{cases},
\end{equation}
where $\sigma_{R_{1/2}}^{x/z}$ are Pauli matrices acting on reference qubits $R_{1}$/$R_{2}$, and $\rho^{0}_{R_{1/2}Q}$ is the maximally entangled Bell state. The projectors $P_{s/p}$ project onto the $+1$ eigenstates of $A_{s}$/$B_{p}$,
\begin{equation}
P_{s}=\frac{1}{2}\left(\mathbb{I}+A_{s}\right),\ P_{p}=\frac{1}{2}\left(\mathbb{I}+B_{p}\right).
\end{equation}
$\rho_Q$ is obtained from $\rho_{RQ}$ by tracing out $R$, yielding
\begin{equation}
\rho_{Q}\equiv(\frac{1}{2}\mathbb{I})\times (\frac{1}{2}\mathbb{I})\times\left(\prod_{\forall s}P_{s}\right)\times\left(\prod_{\forall p}P_{p}\right).\label{eq: rho_Q_TC}
\end{equation}

We now derive $\rho_{RQ^\prime/Q^\prime}$ (see Eq. \eqref{eq:rho_Qprime_toric}), which results from error-corrupted $\rho_{RQ/Q}$. These errors manifest as excitations of $H_{\text{TC}}$,
driving  the system away from its initial state. Error correction
requires to detect these excitations. This is feasible due to the high degree of symmetries
$0=\left[A_{s},\ H_{\text{TC}}\right]=\left[B_{p},\ H_{\text{TC}}\right]$,
allowing one to label all excitations by symmetry charge from $A_{s}$ or $B_{p}$.
However, these excitations are non-local, e.g., string excitations
with negative $A_{s}/B_{p}$ charges at their ends (see Fig.~\ref{fig:toric_code}
(b) for an illustration). This non-locality complicates error syndrome
identification, potentially hindering correction. For example, under
thermal noise, string excitations of varying lengths are equally probable,
making correction inevitably alter the code space---a phenomenon
known as the thermal fragility of the toric code \cite{Nussinov2008prb}. In contrast, local errors have a finite threshold because they differentiate string excitations by assigning string tension. Specifically, we consider the local bit-flip
and phase error channels (denoted by $\mathcal{E}_{X,\ l}$ and $\mathcal{E}_{Z,\ l}$),
giving rise to the following error corrupted density matrix, 
\begin{equation}\label{eq:rho_RQ_p}
\rho_{RQ^{\prime}/Q^\prime}=\circ_{\forall l}\left(\mathcal{E}_{Z,\ l}\circ\mathcal{E}_{X,\ l}\right)\left[\rho_{RQ/Q}\right],
\end{equation}
and, similar to Eq.~\eqref{eq:noise0},
\begin{equation}
%\begin{cases}
\mathcal{E}_{X/Z,\ l}\left[\rho\right]=\left(1-p\right)\ \rho+p\ \sigma_{l}^{x/z}\ \rho\ \sigma_{l}^{x/z}
%\\
%\mathcal{E}_{Z,\ l}\left[\rho_{RQ/Q}\right]=\left(1-p\right)\ \rho_{RQ/Q}+p\ \sigma_{l}^{z}\ \rho_{RQ/Q}\ \sigma_{l}^{z}
%\end{cases}.
\end{equation}
This density matrix $\rho_{RQ^\prime/Q^{\prime}}$ represents an ensemble of
string excitations (or error chains) weighted by the string length \cite{dennis2002jmp}:
\begin{eqnarray}
\rho_{RQ^{\prime}/Q^\prime} & = & \sum_{\left\{ \mathcal{S}_{X}\right\} }\sum_{\left\{ \mathcal{S}_{Z}\right\} }P_{m}\left[\mathcal{S}_{Z}\right]P_{e}\left[\mathcal{S}_{X}\right]\nonumber \\
 &  & \times\mathcal{W}_{m}\left[\mathcal{S}_{Z}\right]\mathcal{W}_{e}\left[\mathcal{S}_{X}\right]\rho_{RQ/Q}\mathcal{W}_{e}\left[\mathcal{S}_{X}\right]\mathcal{W}_{m}\left[\mathcal{S}_{Z}\right],\nonumber\\ \label{eq:rho_Qprime_toric}
\end{eqnarray}
where $\mathcal{S}_{Z}$ ($\mathcal{S}_{X}$) denotes strings on the (dual)
lattice with lengths $\left|\mathcal{S}_{Z}\right|$ $\left(\left|\mathcal{S}_{X}\right|\right)$
(see Fig.~\ref{fig:toric_code} (c)).  The operators $\mathcal{W}_{e}\left[\mathcal{S}_{X}\right]=\prod_{l\in\mathcal{S}_{X}}\sigma_{l}^{x}$ and $\mathcal{W}_{m}\left[\mathcal{S}_{Z}\right]=\prod_{l\in\mathcal{S}_{Z}}\sigma_{l}^{z}$
generate $e$ and $m$ excitations on top of the ground state, respectively, with probabilities
\begin{equation}
\begin{cases}
P_{e}\left[\mathcal{S}_{X}\right]=\left(1-p\right)^{N}\times\left(\frac{p}{1-p}\right)^{\left|\mathcal{S}_{X}\right|}\\
P_{m}\left[\mathcal{S}_{Z}\right]=\left(1-p\right)^{N}\times\left(\frac{p}{1-p}\right)^{\left|\mathcal{S}_{Z}\right|}
\end{cases}.\label{eq:prob_e_m}
\end{equation}
 The exponents $\left|\mathcal{S}_{X}\right|$ and $\left|\mathcal{S}_{Z}\right|$
reflect the string tension $-\ln\left(\frac{p}{1-p}\right)$,
 suppressing longer error chains \cite{fnGibbsToric}.

\subsubsection{Coherent information and the random-bond Ising model}
For the decohered toric code, the R\'enyi-$n$ CI can be represented via the $(n-1)$-flavor RBIM \cite{fan2024prxQ},
\begin{equation}
I_c^{(n)} =-\frac{2}{n-1}\log_{2}\frac{\sum_{\alpha}Z_{\text{RM},\ \alpha}^{\left(n\right)}\left[K\right]}{2^{n-1}Z_{\text{RM}}^{\left(n\right)}\left[K\right]},\ \text{with}\ \ p = \frac{e^{-K}}{2\cosh K}.\label{eq:Renyi_n_CI_2qubit}
\end{equation}
Here, $\alpha = PP, AP, PA, AA$ denotes the four possible combinations of periodic (P) and antiperiodic (A) boundary conditions in $x$ and $y$ directions, for each flavor $f$ of Ising spins on a torus. We will suppress the $PP$ index for the periodic boundary sector, unless stated otherwise.
$Z_{\text{RM},\ \alpha}^{\left(n\right)}$, the partition function for the $(n-1)$-flavor RBIM with $\alpha$-boundary conditions, is
\begin{equation}
\begin{cases}
Z_{\text{RM},\ \alpha}^{\left(n\right)}\left[K\right]\equiv\sum_{\left\{ \sigma=\pm1,\ \eta=\pm1\right\} }P\left[\eta;\ K\right]e^{-H^{(n)}_{\text{RM},\ \alpha}\left[\eta;\ K\right]}\\
H^{(n)}_{\text{RM},\ \alpha}\left[\eta;\ K\right]\equiv-K\sum_{f=1}^{n-1}\sum_{\langle i,\ j\rangle}\eta_{ij}\sigma_{i}^{\left(f\right)}\sigma_{j}^{\left(f\right)}
\end{cases},\label{eq:RBIM_def}
\end{equation}
where $\eta_{ij}=\pm 1$ is a random bond drawn from the distribution
\begin{equation}
P\left[\eta;\ K\right]=\prod_{\langle i,\ j\rangle}\frac{e^{K\eta_{ij}}}{2\cosh\left(K\right)}.
\end{equation}
 Eq.~\eqref{eq:Renyi_n_CI_2qubit} is established via two observations: (a) States in the
code space ($|\Psi_{\text{TC}}\rangle$) consist of loop superpositions,
satisfying $+|\Psi_{\text{TC}}\rangle=A_{s}|\Psi_{\text{TC}}\rangle=B_{p}|\Psi_{\text{TC}}\rangle$.
(b) The terms $\text{Tr}\rho_{Q^{\prime}}^{n}$ and $\text{Tr}\rho_{RQ^{\prime}}^{n}$
measure the overlap of $|\Psi_{\text{TC}}\rangle$ connected by string
excitations, e.g.,
$\langle\Psi_{\text{TC}}^{\left(f+1\right)}|\mathcal{W}_{e/m}^{\left(f+1\right)}
  \times  \mathcal{W}_{e/m}^{\left(f\right)}|\Psi_{\text{TC}}^{\left(f\right)}\rangle$,
where $f\in\left[1,\ n\right]$ labels states/operators from the $f$-th
density matrix within $n$ replicas, with $f=n+1$ identified as $f=1$. Hence, both $\text{Tr}\rho_{Q^{\prime}}^{n}$
and $\text{Tr}\rho_{RQ^{\prime}}^{n}$ are non-zero only when the error chain $\mathcal{S}_{X/Z}^{\left(f+1\right)}-\mathcal{S}_{X/Z}^{\left(f\right)}$
forms a closed loop (see Fig. \ref{fig:toric_code} (d)). Using the
$f=n$ strings as a reference, $\text{Tr}\rho_{Q^{\prime}}^{n}$ and
$\text{Tr}\rho_{RQ^{\prime}}^{n}$ equal the sum of $\left(\frac{p}{1-p}\right)^{\left|\mathcal{C}_{Z/X}^{\left(f\right)}\right|}$
for all possible loop configurations $\mathcal{C}_{Z/X}^{\left(f\right)}|_{f\neq n}\equiv\mathcal{S}_{Z/X}^{\left(f\right)}-\mathcal{S}_{Z/X}^{\left(n\right)}$. This sum coincides with the loop expansion of the RBIM, 
with the random bond $\eta_{ij}$ from the reference string configuration
at $f=n$, subjected to $P\left[\eta;\ K\right]=\prod_{\langle i,\ j\rangle}\frac{e^{K\eta_{ij}}}{2\cosh\left(K\right)}$.
A key difference between $\text{Tr}\rho_{Q^{\prime}}$ and $\text{Tr}\rho_{RQ^{\prime}}$
arises from the reference qubit: For $\text{Tr}\rho_{RQ^{\prime}}^{n}$, $\mathcal{S}_{X/Z}^{\left(f+1\right)}-\mathcal{S}_{X/Z}^{\left(f\right)}$
must form a contractible loop, as purification by $R$ implies a one-dimensional
space spanned by $|\Psi_{\text{TC}}\rangle$, while $\text{Tr}\rho_{Q^{\prime}}^{n}$
involves non-contractible loops, leading
to tunneling between different states in the code space, which is
reflected as different boundary conditions in the RBIM.

\begin{figure}
\includegraphics[scale=0.18]{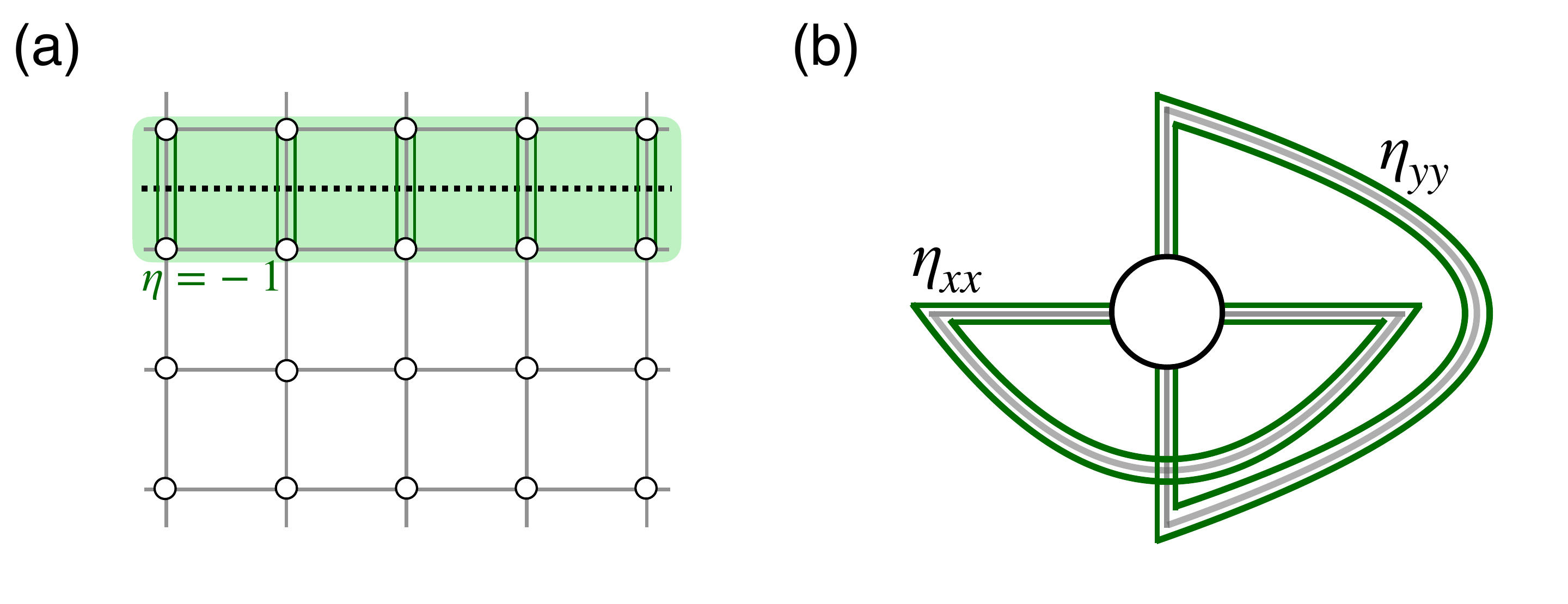}
\caption{Illustration of (a) the equivalence of anti-periodic boundary conditions with a $\mathbb{Z}_2$ flux line insertion, and (b) the random-bond Ising model (RBIM) in the one-site limit. In (a), an anti-periodic boundary along the $y$-axis is equivalent to a $\mathbb{Z}_2$ flux line ($\eta=-1$) along the $x$-axis, resulting in $\prod_{\langle i,\ j\rangle\in \mathcal{L}_y}\eta_{ij}=-1$ for any non-contractible loop in the $y$ direction. In (b), the hollow circle represents the Ising spin, with the $\mathbb{Z}_2$ flux lines $\eta_{xx}$ and $\eta_{yy}$ changing the interaction sign (green double line).\label{fig:Ising_1site}}
\end{figure}

The RBIM features a $\mathbb{Z}_2$ gauge symmetry, with $\eta_{ij}$ as the $\mathbb{Z}_2$ gauge field \cite{fradkin1978prb}, defined by
\begin{equation}
\sigma_{i}^{(f)}\rightarrow \sigma_{i}^{(f)}\tau_i,\ \eta_{ij}\rightarrow \tau_i  \eta_{ij}\tau_j,\ \text{and}\ \tau_i =\pm1,
\end{equation} 
which leaves $H_{\text{RM},\ \alpha}^{(n)}$ invariant. The CI is thus tied to the free energy of the $\mathbb{Z}_2$ flux line, as the $\alpha$-boundary condition corresponds to inserting such a flux line  (see Fig.~\ref{fig:Ising_1site} (a) for an illustration).
This connection becomes particularly transparent in the single-site limit of the RBIM (see Fig.~\ref{fig:Ising_1site} (b)), where the code space corresponds to different boundary condition sectors. In this limit, $I_c^{(n)}$ becomes
\begin{equation}
I_c^{(n)}=-\frac{2}{n-1}\log_2\left\{\sum_{\eta} P[\eta]\left[\frac{e^{K(\eta_{xx}+\eta_{yy})}}{(2\cosh K)^2}\right]^{(n-1)}\right\} +2,
\end{equation}
using 
\begin{equation}
\begin{cases}
Z_{\text{RM}}^{(n)}[K]\overset{N=1}{\implies}\sum_{\{\eta\}}P\left[\eta\right]e^{\left(n-1\right)K\left(\eta_{xx}+\eta_{yy}\right)},\\
\sum_{\alpha}Z_{\text{RM},\ \alpha}^{(n)}[K]\overset{N=1}{\implies}\left(2\cosh K\right)^{2\left(n-1\right)}
\end{cases}.
\end{equation}
Here, $\eta_{xx}, \eta_{yy} = \pm1$ function both as $\mathbb{Z}_2$ flux lines and as different boundary conditions. Thus, we confirm that $I_c^{(n)}$ represents the free energy of the $\mathbb{Z}_2$ flux lines, since $\frac{e^{K(\eta_{xx} + \eta_{yy})}}{(2 \cosh K)^2}$ is the corresponding Boltzmann weight. Additionally, this expression matches the R\'enyi-$n$ CI for two decoupled qubits (Eq.~\eqref{eq:Renyi_n_CI_2qubit}) with $p = \frac{e^{-K}}{2 \cosh K}$, indicating that the code space consists of different boundary condition sectors.

\subsubsection{Numerical results for coherent information \label{subsubsec:numres_DeTC}} 

\begin{figure}%[h!]
\includegraphics[scale=0.18]{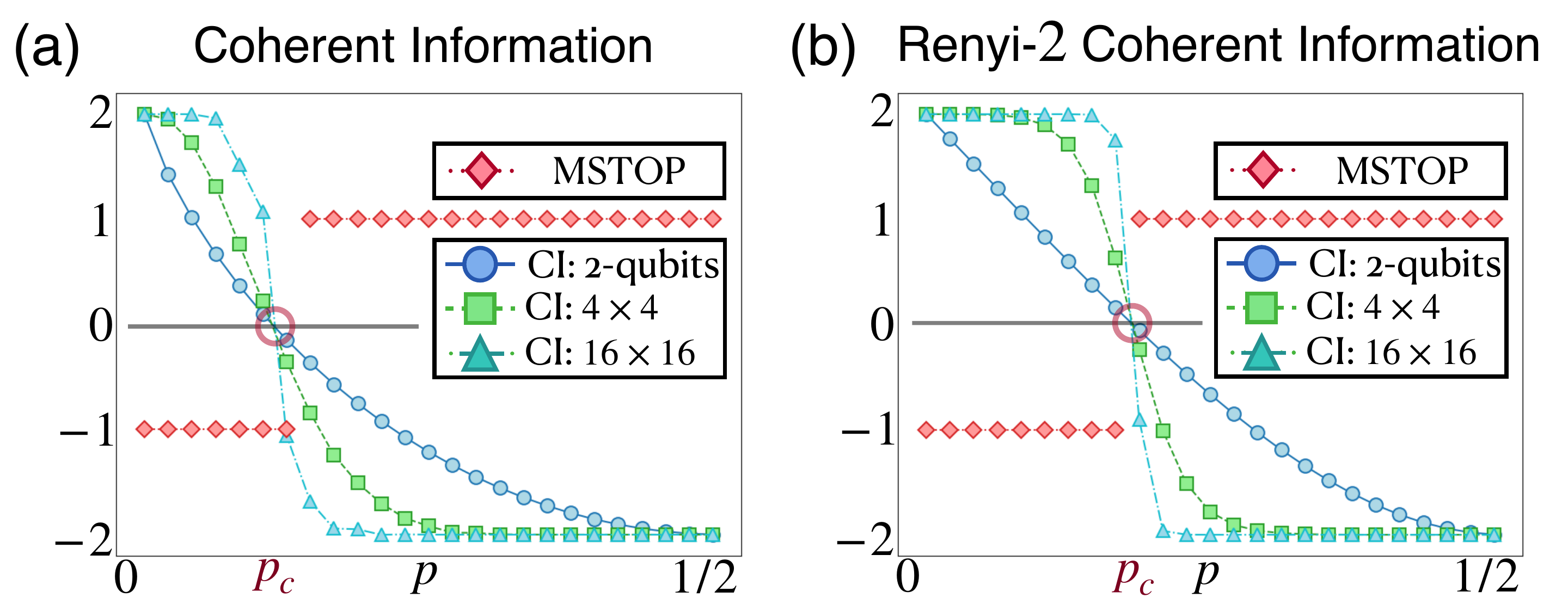}

\caption{Numerical results for the coherent information (CI) (a) and R\'enyi-$2$
CI (b) in the decohered toric code, calculated using the random-bond Ising model (RBIM) partition function. The crossing
point of these two quantities at different system sizes
shows only tiny finite-size effects, marking the critical point of the
decoding phase transition. In the alternative representation as a class D Majorana (Sec. \ref{sec:TSC}), this point aligns with the phase boundary identified by the mixed-state topological order parameter (MSTOP), defined as $\text{sign}\left[\prod_\alpha\langle (-)^{\hat{Q}}\rangle_\alpha \right]$
where `$\alpha$' represents periodic/antiperiodic
boundary conditions, and $\left(-1\right)^{\hat{Q}}$ denotes the fermion
parity. \label{fig:CI_Renyi2}}
\end{figure}

Fig. \ref{fig:CI_Renyi2} presents numerical results for the CI and its R\'enyi-$2$ counterpart. The data shows that the CI for different system sizes crosses at a common point with only tiny finite-size drift, reproducing the optimal error threshold.
This phenomenon extends beyond the toric code, and has been observed
in various code models \cite{colmenarez2024prr}, e.g., planar codes, color codes and quantum low-density parity check codes. We will rationalize this from two perspectives, duality in the RBIM (Sec. \ref{sec:dToric_CI}) and via the topological phase transition in disordered Majorana fermions (Sec. \ref{sec:TSC}).

\section{Coherent information and duality \label{sec:dToric_CI}}

We now link CI to the low/high temperature duality (see e.g., Ref. \cite{shankar2017cambridge}).
Our main result here is a duality relation between the $n-1$ flavor random-bond Ising model (RBIM) on a torus ($Z_{\text{RM}}^{(n)}$) and its dual
($\tilde{Z}_{\text{RM}}^{(n)}$), which holds for an \emph{arbitrary} number of sites $N$,
\begin{equation}
2^{n-1}\tilde{Z}_{\text{RM}}^{\left(n\right)}\left[K\right]=\sum_{\left\{ \alpha\right\} }Z^{(n)}_{\text{RM,\ \ensuremath{\alpha}}}\left[K\right].\label{eq:dual_nRBIM}
\end{equation}
This involves distinct boundary-condition sectors in the dual model, setting it apart from previous results (e.g., \cite{nishimori2002jpsj,nishimori2007jsp, ohzeki2009pre}), which focus solely on the thermodynamic limit. At the same time, this is a crucial stepping stone for the mapping to fermions established below.

\begin{figure}
\includegraphics[scale=0.18]{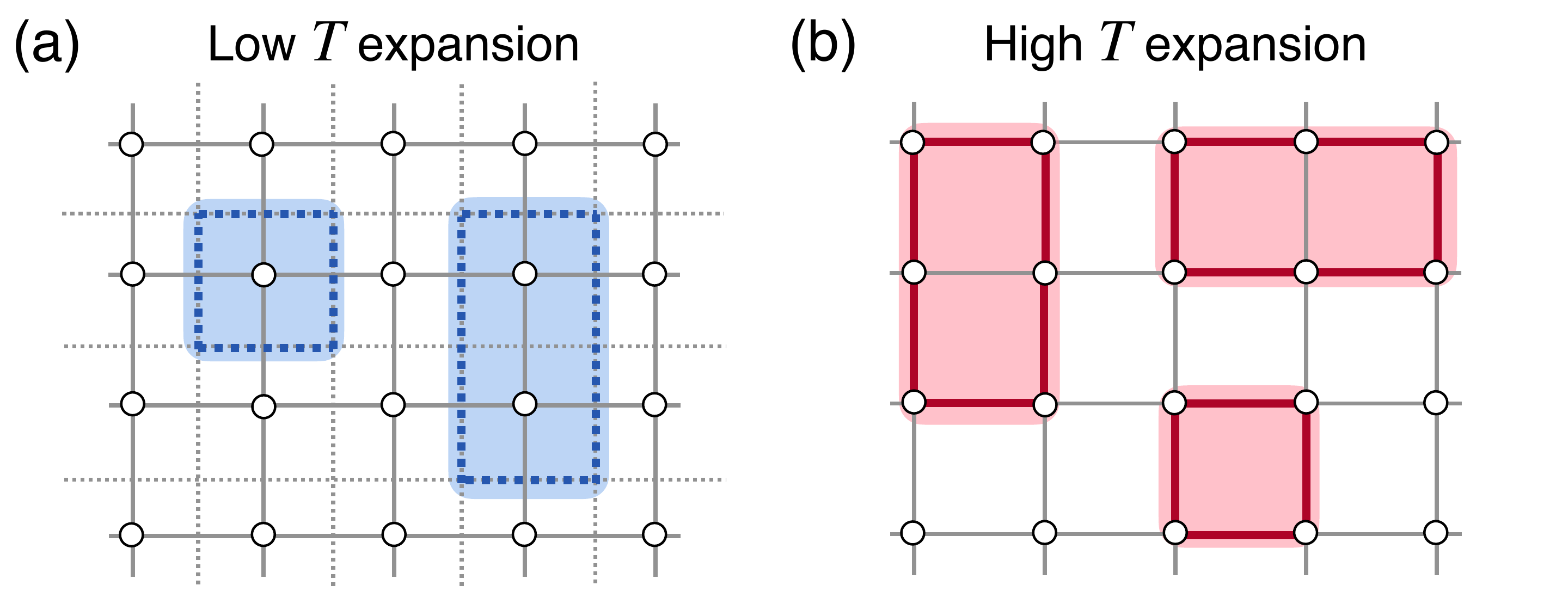}\caption{Low-temperature  (a) and high temperature expansion (b) in
the Ising model on a square lattice: The partition function for the
Ising model is $Z_{\text{IM}}[J]=\sum_{\left\{ \sigma\right\} }e^{\sum_{\langle i,\ j\rangle}J\sigma_{i}\sigma_{j}}$
, where the Ising spins are represented by hollow circles, and the
grey dashed lines indicate the dual lattice. In the low-temperature
limit $(J\gg1)$, this partition function contains domain-wall excitations
above the ground state (indicated by the blue dashed line in (a)), with energy
proportional to the domain wall perimeter, $2J\left|\mathcal{C}\right|$.
In the high temperature limit ($J\ll1$), the partition function is
expressed as a sum over different loop configurations due to the expansion (red solid line in (b)):
$Z_{\text{IM}}=\left(\cosh J\right)^{2N}\sum_{\left\{ \sigma\right\} }\prod_{\langle i,\ j\rangle}\left(1+\tanh J\sigma_{i}\sigma_{j}\right)$.
\label{fig:lT_Ising}}
\end{figure}

The boundary conditions are crucial for error correction, as they encode logical information, as previously shown. The occurrence of different boundary condition sectors affords a simple picture: 
In the low temperature regime ($K\gg 1$), $Z_{\text{RM}}^{(n)}\left[K\right]$
features domain wall excitations over $2^{n-1}$ degenerate ground states
(all spins up/down for each flavor) (see Fig. \ref{fig:lT_Ising} (a) for an illustration via the Ising model).
On the dual lattice, these domain walls corresponds to \textit{contractible}
loop configurations. Conversely, the high-temperature expansion of
$\tilde{Z}_{\text{RM}}^{(n)}\left[K\right]$ includes all loop excitations,
including \textit{\textcolor{black}{non-contractible}} ones (see Fig.
\ref{fig:lT_Ising} (b)). The mismatch in loop configurations is captured by different boundary conditions in the high and low temperature expansions, with an extra factor of $2^{n-1}$ accounting
for the degenerate ground states. Finally, in the $n=2$ case, we recover the self-dual relation for the Ising model ($Z_{\text{IM}, \alpha}$) \cite{wegner2014arxiv},
\begin{equation}
2 \tilde{Z}_{\text{IM}}[J] = \sum_{\alpha} Z_{\text{IM}, \alpha}[J], 
\end{equation}
where $Z_{\text{IM}, \alpha}[J] = \sum_{\{\sigma\}}e^{J\sum_{\langle i,\ j\rangle}\sigma_i \sigma_j}$, and it corresponds to the R\'enyi-$2$ RBIM $Z_{\text{RM}, \alpha}^{(2)}[K]$ via $\tanh J = (\tanh K)^2$ (see Sec. \ref{subsec:Renyi_2_CI_Ising} for details).

Equation~\eqref{eq:dual_nRBIM} connects the R\'enyi-$n$ CI zero point to the self-dual point. Applying it to Eq. \eqref{eq:Renyi_n_CI_2qubit}, we find
\begin{equation}
I_{c}^{\left(n\right)}=-\frac{2}{n-1}\log_{2}\frac{\tilde{Z}_{\text{RM}}^{\left(n\right)}\left[K\right]}{Z_{\text{RM}}^{\left(n\right)}\left[K\right]},
\end{equation}
confirming the coincidence of the self-dual point ($K=K_{\text{SD}}$) and $I_c^{(n)}=0$,
\begin{equation}
K=K_{\text{SD}}:\  Z_{\text{RM}}^{(n)}[K_{\text{SD}}]=\tilde{Z}_{\text{RM}}^{(n)}[K_{\text{SD}}]\iff I_c^{(n)} =0.
\end{equation}
Assuming (a) only two phases and (b) identical forms for the model and its dual, the self-dual point coincides with the critical point, as argued by Kramers and Wannier: In the thermodynamic limit, the partition function's singularity occurs solely at the critical point, which is necessarily at the self-dual point. However, this exact alignment between the self-dual and critical points holds only for certain replica indices (e.g., $n=2,\ 3,\ \infty$) as assumption (b) may not hold for other indices.

We will now quantitatively define the dual model (i.e., $\tilde{Z}_{\text{RM}}^{(n)}$ in Eq. \eqref{eq:dual_nRBIM_def} and $\tilde{Z}_{\text{IM}}$ in Eq. \eqref{eq:dual_Ising}), and derive the duality relation Eq. \eqref{eq:dual_nRBIM} using the Wu-Wang method \cite{wu1976jmp}. This method
provides a systematic approach for duality transformations by connecting the low/high-temperature expansions
with a Fourier transformation. We begin with the R\'enyi-$2$ CI represented by $Z_{\text{RM}}^{\left(2\right)}$, which reduces to the clean Ising model (all bond variables $\eta$ positive), and is analytically solvable. We then take the limit $n\rightarrow1^{+}$. 

\subsection{Duality transformation from Fourier transformation in the Ising model}

\begin{figure}
\includegraphics[scale=0.18]{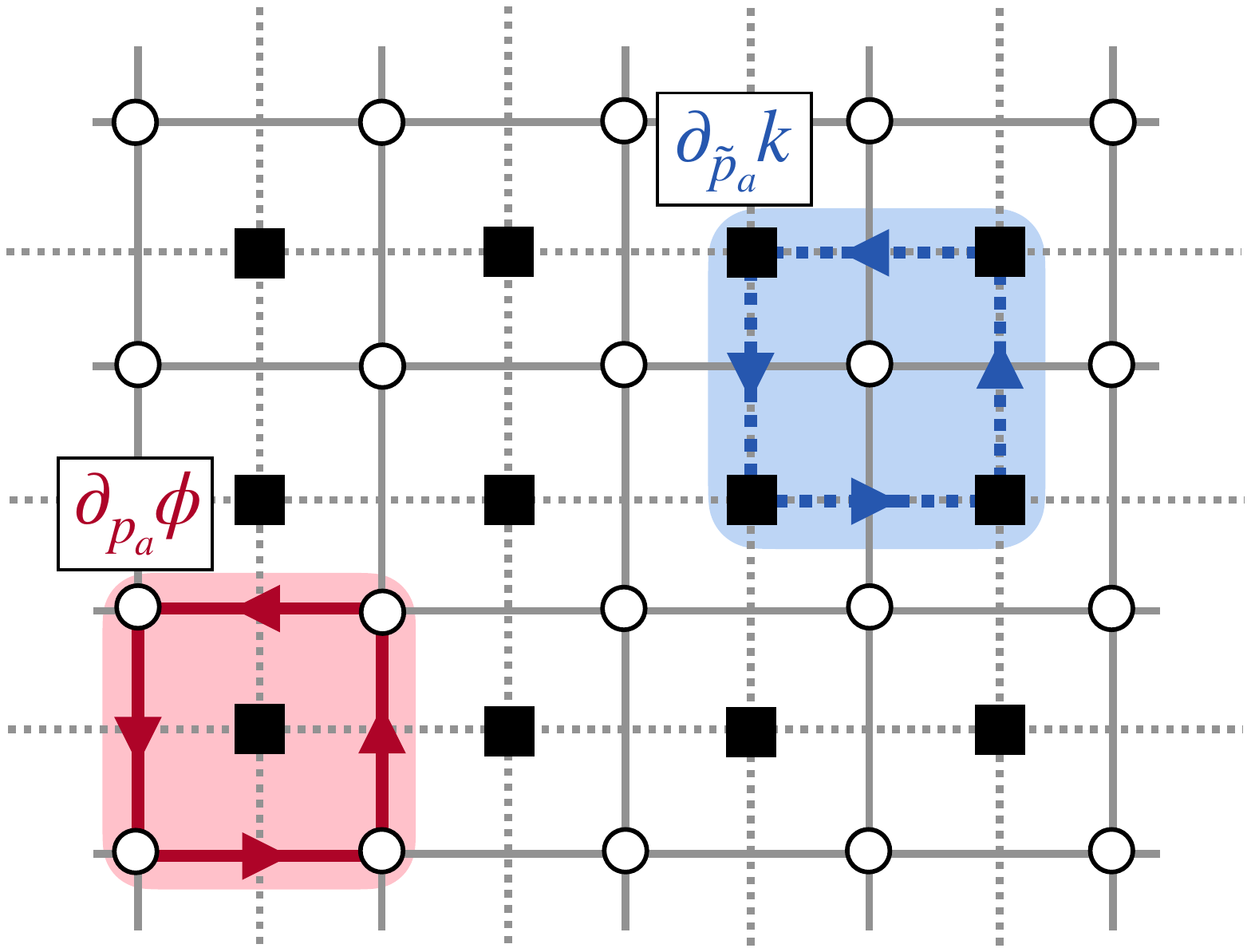}\caption{Definition of curl $\partial_{p_a}\phi$ ($\partial_{\tilde{p}_a}k$) on the (dual) lattice, with hollow circles (solid squares) for (dual) Ising spins. Arrows indicate the sign of $\phi_{ij}$ ($k_{ij}$): positive for right/up, negative otherwise.
 \label{fig:curl_lattice_def}}
\end{figure}

For clarity, we explain the Wu-Wang method -- and introduce the definition of the dual model --  using the clean Ising model. Furthermore, the Ising model is closely related to the case $n=2$, see Eq.~\eqref{eq:Ising-RM} below. A key point here is that this method, while previously used in the thermodynamic limit only, is applicable to  finite-size systems, and allows us to keep track of the boundary conditions, in turn encoding the quantum information content. The Ising partition
function is 
\begin{equation}\label{eq:Ising_link}
Z_{\text{IM},\ \alpha}\left[J\right]=\sum_{\left\{ \sigma\right\} }e^{J\sum_{\langle i,\ j\rangle}\sigma_{i}\sigma_{j}}=\sum_{\left\{ \sigma\right\} }\prod_{\langle i,\ j\rangle}x_{ij,\ \alpha}\left[J;\ \phi\right],
\end{equation} where $x_{ij,\ \alpha}\left[J\right]=e^{J\sigma_{i}\sigma_{j}}$ represents
interactions on the bond $\langle i,\ j\rangle$ for Ising spin $\sigma_i$ with $\alpha$-boundary condition. For later convenience,
we introduce the bond variable $\phi_{ij}$, 
\begin{equation}
\cos\left(\pi\phi_{ij}\right)\equiv\sigma_{i}\sigma_{j}\implies x_{ij,\ \alpha}\left[J;\ \phi\right]=e^{J\cos\left(\pi\phi_{ij}\right)}.
\end{equation}
We start by defining the dual model for the partition function under periodic boundary conditions ($Z_{\text{IM}}[J]$), and then obtain the other cases by imposing twisted boundary conditions on the Ising spins.  Considering $Z_{\text{IM}}[J]$ in the low-temperature phase ($J\gg1$), $\phi_{ij}$ prefers homogeneous configurations,
while it fluctuates strongly in the high-temperature limit, such that the partition sum converges slowly in this regime. This motivates
defining the dual model $\tilde{Z}_{\text{IM}}[J]$,
which describes the opposite temperature limit with rapid convergence. It involves 
variables $\tilde{x}_{ij}$ obtained from $x_{ij}\left[\phi\right]$
via Fourier transformation, 
\begin{eqnarray}
\tilde{Z}_{\text{IM}}\left[J\right]&\equiv&2\times\sum_{\left\{ k\right\} }\prod_{\langle i,\ j\rangle}\tilde{x}_{ij}\left[J;\ k\right]\nonumber\\
{}&=&\left(\frac{e^J}{\sqrt{2}\cosh \tilde{J}}\right)^{2N} \sum_{\{\tilde{\sigma}\}} e^{\tilde J \sum_{\langle i,\ j\rangle}\tilde{\sigma}_i \tilde{\sigma}_j}.
\label{eq:dual_Ising}
\end{eqnarray}
In the second line, we introduce Ising spins $\tilde{\sigma}_i=\pm 1$ residing at the dual lattice (solid square in Fig. \ref{fig:curl_lattice_def}), such that $e^{i\pi k_{ij}}=\tilde{\sigma}_i \tilde{\sigma}_j$. The factor 2 in front of $\sum_{\{k\}}$ accounts for the global $\mathbb{Z}_2$ symmetry of the Ising spins, ensuring $2\sum_{\{k\}}=\sum_{\{\tilde{\sigma}\}}$.  The  variable $\tilde{x}_{ij}$ is defined as
\begin{eqnarray}
\tilde{x}_{ij}\left[J;\ k\right] & \equiv & \frac{1}{\sqrt{2}}\sum_{\phi_{ij}=0,1}e^{i\pi k_{ij}\phi_{ij}}x_{ij}\left[J;\ \phi\right]\nonumber \\
 & = & \frac{1}{\sqrt{2}}\frac{e^{J}}{\cosh \tilde{J}}x_{ij}\left[\tilde{J};\ k\right],\label{eq:Dual_ising_boltzmann_weight}
\end{eqnarray}
with $\sum_{\phi_{ij}=0,1}$ treating $\phi_{ij}$ at different bonds independently and $\tanh \tilde{J}=e^{-2J}$, which expresses the high/low temperature duality between $Z_{\text{IM}}[J]$ and $\tilde{Z}_{\text{IM}}[J]$. For other sectors $\tilde{Z}_{\text{IM},\ \alpha}[J]$, the  variable follows the same form as $x_{ij}$, but with its Ising spins subjected to different boundary conditions.

We now aim to establish an exact relation between $Z_{\text{IM}}$
and its dual, i.e., 
\begin{equation}
2\tilde{Z}_{\text{IM}}\left[J\right]=\sum_{\alpha}Z_{\text{IM},\ \alpha}\left[J\right],\label{eq:dual_model}
\end{equation}
where the different boundary conditions arise from the torus topology, as explained below. We derive this result by expressing the bond variable $\tilde{x}_{ij}[J;\ k]$ of $\tilde{Z}_{\text{RM}}^{(n)}$ in terms of $x_{ij}[J;\ \phi]$ and integrating out $k_{ij}$. Detailed derivations are provided in Appendix \ref{sup_sec_Dirac_delta}, and we concentrate here on the key steps. Using the definition of $\tilde{Z}_{\text{IM}}$ (Eq. \eqref{eq:dual_Ising}), we have
\begin{equation} \label{eq:Dirac_delta_derivation}
\tilde{Z}_{\text{IM}}\left[J\right]=2\sum_{\left\{k,\ \phi\right\},\ \partial k=0\,\text{mod}\,2} \prod_{\langle i,\ j\rangle}\left[\frac{1}{\sqrt{2}}e^{i\pi k_{ij}\phi_{ij}}x_{ij}\left[J;\ \phi\right]\right].
\end{equation}
The $k_{ij}$ variables are subject to the curl-free constraint $\partial k = 0 \ \text{mod}\ 2$ for all contractible and non-contractible loops on the dual lattice (see blue dashed lines in Fig. \ref{fig:curl_lattice_def} for $\partial_{\tilde{p}_a}k$ associated with a dual plaquette). This constraint follows from the definition $\tilde{\sigma}_i\tilde{\sigma}_j=e^{i\pi k_{ij}}$, implying $\prod_{\langle i,\ j\rangle\in \tilde{\mathcal{C}}}e^{i\pi k_{ij}}=1$ for any loop $\tilde{\mathcal{C}}$ on the dual lattice with 'mod 2' arising from the $2\pi$ ambiguity in the exponent. The 'mod 2' condition is implicit for all curls (e.g., $\partial_{\tilde{p}_a} k$) and will be omitted hereafter.
Summing over $k_{ij}$ results in a product of Dirac-delta functions,
\begin{equation}
2\sum_{\left\{ k\right\},\ \partial k =0 }\ \prod_{\langle i,\ j\rangle}e^{i\pi k_{ij}\phi_{ij}}=\frac{1}{2}\prod_{p_{a}}\left[2\delta\left(\partial_{p_{a}}\phi\right)\right].\label{supp_eq:Dirac_delta}
\end{equation}
 Here, $\partial_{p_a} \phi$ represents the curl for plaquette $p_a$ (red solid line in Fig. \ref{fig:curl_lattice_def}). 
 The corresponding Dirac-delta functions $\delta(\partial_{p_a}\phi)$ impose in turn curl-free constraints on $\phi_{ij}$, which can be solved by introducing variables $\phi_i = 0, 1$ on lattice sites, ensuring $\partial_{p_a} \phi_{ij} = 0$, i.e., $\phi_{ij} = \phi_i - \phi_j$.
However, on a torus, non-trivial homology requires four boundary condition sectors beyond this local solution. Specifically, the number of independent $\phi_{ij}$ variables is $\#\phi_{ij} = \frac{2^{2N}}{2^{N-1}}$, with $2^{2N}$ from the bonds and $2^{N-1}$ from the Dirac-delta constraints. The number of independent $\phi_i$ variables is $\#\phi_i = \frac{2^{N}}{2}$, where the factor of $\frac{1}{2}$ accounts for the global shift symmetry ($\phi_i \rightarrow \phi_i + 1$). The mismatch between $\phi_i$ and $\phi_{ij}$ is thus $\frac{\#\phi_i}{\#\phi_{ij}} = \frac{2^{N-1}}{2^{2N}/2^{N-1}} = \frac{1}{4}$, reflecting the four distinct boundary condition sectors associated with the torus topology.

An intriguing aspect of the Ising model is that $Z_{\text{IM}}$
and its dual $\tilde{Z}_{\text{IM}}$ share the same form (i.e.,
$\tilde{x}_{ij}\left[J\right]\propto e^{\tilde{J}\tilde{\sigma_{i}}\tilde{\sigma}_{j}}$). 
Based on this and assuming only two phases exist, Kramers and Wannier
found that the critical point ($J=J_{c}$) coincides with the self-dual
point ($J=J_{\text{SD}}$), as the partition function has a singularity only at this point in the thermodynamic limit, 
\begin{equation}
Z_{\text{IM}}\left[J_{\text{SD}}\right]=\tilde{Z}_{\text{IM}}\left[J_{\text{SD}}\right].\label{eq:self_dual_Ising}
\end{equation}
This condition holds when $J=\tilde{J}=J_{\text{SD}}$ for any system size, thereby determining the location of the critical point in the thermodynamic limit,
\begin{equation}
N\rightarrow \infty:\ \tanh J_{c}=\tanh J_{\text{SD}}=\sqrt{2}-1.
\end{equation}

\subsection{R\'enyi-$2$ Coherent information and Kramers-Wannier duality in the Ising
model }\label{subsec:Renyi_2_CI_Ising}

Via the self-duality relation Eqs. \eqref{eq:dual_model} and \eqref{eq:self_dual_Ising}, we identify that the self-dual point coincides with the critical point, as it is the only point that can exhibit a free energy singularity under the assumption of two phases. Consequently, the zero point of the R\'enyi-2 CI detects the critical point in the thermodynamic limit, due to an exact relation between R\'enyi-2 CI and self-duality implied by Eqs.~(\ref{eq:dual_model},\ref{eq:self_dual_Ising}),
\begin{equation}
I_{c}^{\left(2\right)}=2\log_{2}\frac{2Z_{\text{IM}}\left[J\right]}{\sum_{\alpha}Z_{\text{IM},\ \alpha}\left[J\right]}=-2\log_{2}\frac{\tilde{Z}_{\text{IM}}\left[J\right]}{Z_{\text{IM}}\left[J\right]},\label{eq:Renyi_2_Ising}
\end{equation}
with
\begin{equation}
\tanh J = (1-2p)^2,
\end{equation}
which implies that 
\begin{equation}
 I_{c}^{\left(2\right)}|_{J=J_{\text{SD}}}=0,
\end{equation}
and thus in the thermodynamic limit, the critical point locates at
\begin{equation}
N\rightarrow\infty:\ \begin{cases}
 J_{c}= J_{\text{SD}}=\text{arctanh}(\sqrt{2}-1)\\
p_{c}=\frac{1}{2}\left(1-\sqrt{\sqrt{2}-1}\right).
\end{cases}
\end{equation}
Here, $J_{\text{SD}}$ is independent of system size, and $p_c$ is the critical error rate (see Fig.
\ref{fig:CI_Renyi2} (b) for numerical results). 

Quantitatively, the
connection between the Ising model and the $Z_{\text{RM}}^{\left(2\right)}\left[K\right]$
is
\begin{equation}\label{eq:Ising-RM}
Z_{\text{RM}}^{\left(2\right)}\left[K\right]=\left(\frac{\cosh K}{\cosh J}\right)^{2N}Z_{\text{IM}}\left[J\right],
\end{equation}
and 
\begin{equation}
\left(\tanh K\right)^{2}=\tanh J,
\end{equation}
 which follows from integrating out the random bond variables:
\begin{eqnarray}
Z_{\text{RM}}^{\left(2\right)}\left[K\right] & = & \sum_{\left\{ \sigma\right\} }\prod_{\langle i,\ j\rangle}\left[\sum_{\eta_{ij}}\left(\frac{e^{K\eta_{ij}}}{2\cosh K}e^{K\eta_{ij}\sigma_{i}\sigma_{j}}\right)\right]\nonumber \\
 & = & \sum_{\left\{ \sigma\right\} }\prod_{\langle i,\ j\rangle}\left(\cosh K\right)\left[1+\left(\tanh K\right)^{2}\sigma_{i}\sigma_{j}\right]\nonumber \\
 & = & \left(\frac{\cosh K}{\cosh J}\right)^{2N}\sum_{\left\{ \sigma\right\} }e^{J\sum_{\langle i,\ j\rangle}\sigma_{i}\sigma_{j}}.
\end{eqnarray}

\subsection{Duality for the random-bond Ising model}\label{sec:selfdu}
Unlike the Ising model, the RBIM $Z_{\text{RM}}^{(n)}$ generally differs from its dual $\tilde{Z}_{\text{RM}}^{(n)}$, so the Kramers-Wannier argument does not apply. This suggests that the self-dual point and the critical point may be unrelated.
Nishimori conjectured that the self-dual point coincides with the critical point based on numerical results \cite{nishimori2007jsp}. We will demonstrate the emergence of this relation. Our argument has two independent ingredients: First, in this section we show that the zero crossing point of the CI coincides with the self-dual point, for any system size. Second, we show in Sec.~\ref{sec:TSC} in the   Majorana representation that the zero crossing point of the CI coincides with the critical point in the thermodynamic limit. From this we can conclude that the self-dual and the critical point must coincide for extensive systems. 

The formula linking CI and duality is: 
\begin{equation}
I_{c}^{\left(n\right)}=-\frac{2}{n-1}\log_{2}\frac{\tilde{Z}_{\text{RM}}^{\left(n\right)}\left[K\right]}{Z_{\text{RM}}^{\left(n\right)}\left[K\right]},\label{eq:I_c_dual}
\end{equation}
confirming that the self-dual point (i.e., $K=K_{\text{SD}}$:  $\tilde{Z}_{\text{RM}}^{\left(n\right)}=Z_{\text{RM}}^{\left(n\right)}$) coincides with the zero-CI point.
This follows by applying the following duality relation, which generalizes Eq.~\eqref{eq:dual_model}, to Eq. \eqref{eq:Renyi_n_CI_2qubit},
\begin{equation}
2^{n-1}\tilde{Z}_{\text{RM}}^{\left(n\right)}\left[K\right]=\sum_{\left\{ \alpha\right\} }Z^{(n)}_{\text{RM,\ \ensuremath{\alpha}}}\left[K\right].\label{eq: duality_RBIM_n}
\end{equation}
Here, $2^{n-1}$ represents the ground state degeneracy, reducing to 2 for the Ising model ($n = 2$). The right-hand side term includes $4^{n-1}$ boundary condition sectors, accounting for the mismatch in the number of degrees of freedom, $\frac{\# \phi_i^{(f)}}{\# \phi_{ij}^{(f)}} =\left[\frac{2^{N-1}}{2^{2N}/2^{N-1}}\right]^{(n-1)}=\left(\frac{1}{4}\right)^{n-1}$ for $(n-1)$ Ising spin species.

We now derive the duality relation Eq. \eqref{eq: duality_RBIM_n}, following the approach used for the Ising model.  We start by obtaining the expression for the dual model $\tilde{Z}_{\text{RM},\ \alpha}^{\left(n\right)}$, and then derive the  relation by representing its bond variables in terms of those in $Z_{\text{RM}}^{(n)}$. The dual model is given as
\begin{multline}
\tilde{Z}_{\text{RM},\ \alpha}^{\left(n\right)}\left[K\right]\\=\mathcal{N}^{\left(n\right)}\sum_{\left\{ \tilde{\sigma}\right\} }e^{\tilde{K}\sum_{\langle i,\ j\rangle}\left(\sum_{f=1}^{n-1}\tilde{\sigma}_{i}^{\left(f\right)}\tilde{\sigma}_{j}^{\left(f\right)}+\prod_{f=1}^{n-1}\tilde{\sigma}_{i}^{\left(f\right)}\tilde{\sigma}_{j}^{\left(f\right)}\right)}.\label{eq:dual_nRBIM_def}
\end{multline}
Here, $\tilde{\sigma}_{i}^{\left(f\right)}$ represents the $f$-th Ising
spin at site $i$ in the dual lattice, and 
\begin{equation}
\tilde{K}=-\frac{1}{2}\ln\tanh K.
\end{equation}
The prefactor
is given by 
\begin{equation}
\mathcal{N}^{\left(n\right)}\equiv\left(\frac{e^{K}}{\sqrt{2}\cosh \tilde{K}}\right)^{2N\left(n-1\right)}e^{-2N\tilde{K}}.
\end{equation}
For $n=2$, this reproduces the Ising model, confirming its self-dual nature.
Specifically, we derive Eq. \eqref{eq:dual_nRBIM_def} by applying a Fourier transformation to the bond variables. Parallel to the Ising model (Eq. \eqref{eq:Ising_link} and Eq. \eqref{eq:dual_Ising}), we start from the periodic boundary condition sectors, and represent $Z_{\text{RM}}^{(n)}$ and $\tilde{Z}_{\text{RM}}^{(n)}$ as a product of bond variables,
\begin{equation}
\begin{cases}
Z_{\text{RM}}^{\left(n\right)}\left[K\right]=2^{n-1}\sum_{\left\{ \phi^{(f)}\right\} }\prod_{\langle i,\ j\rangle}x_{\text{RM},\ ij}^{(n)}\left[K;\ \phi^{\left(f\right)}\right]\\
\tilde{Z}_{\text{RM}}^{\left(n\right)}\left[K\right]=2^{n-1}\sum_{\left\{ k^{(f)}\right\} }\prod_{\langle i,\ j\rangle}\tilde{x}_{\text{RM},\ ij}^{(n)}\left[K;\ k^{\left(f\right)}\right]
\end{cases}.
\end{equation}
Here, the bond variable $x^{(n)}_{\text{RBIM},\ ij}$ is
\begin{equation}
x_{\text{RM},\ ij}^{\left(n\right)}\left[K;\ \phi^{\left(f\right)}\right]=\sum_{\eta}\frac{e^{K\eta_{ij}}}{2\cosh K}e^{K\sum_{f=1}^{n-1}\cos\left(\pi\phi_{ij}^{\left(f\right)}\right)},
\end{equation}
while its dual variable $\tilde{x}_{\text{RM},\ ij}^{(n)}$, a Fourier
transform of $x_{\text{RM},\ ij}^{\left(n\right)}$, is given by
\begin{multline}
\tilde{x}_{\text{RM},\ ij}^{\left(n\right)}\left[K;\ k^{\left(f\right)}\right] = \left[\left(\frac{e^{K}}{\sqrt{2}\cosh \tilde{K}}\right)^{\left(n-1\right)}e^{-\tilde{K}}\right] \\
 \times e^{\tilde{K}\sum_{f=1}^{n-1}\cos\left(\pi k_{ij}^{\left(f\right)}\right)}e^{\tilde{K}\cos\left(\pi\sum_{f=1}^{n-1} k_{ij}^{\left(f\right)}\right)}.
\end{multline}
This reproduces Eq. \eqref{eq:dual_nRBIM_def} with boundary conditions according to the $\alpha$ index. Finally, we reproduce the duality relation Eq.~\eqref{eq: duality_RBIM_n} by rewriting the bond variable $\tilde{x}_{\text{RM},\ ij}^{(n)}[K;\ k^{(f)}]$ in terms of $x_{\text{RM},\ ij}^{(n)}[K;\ \phi^{(f)}]$ and summing over $k_{ij}^{(f)}$, analogously to the Ising model,  Eq.~\eqref{supp_eq:Dirac_delta}.

\begin{figure}
\includegraphics[scale=0.18]{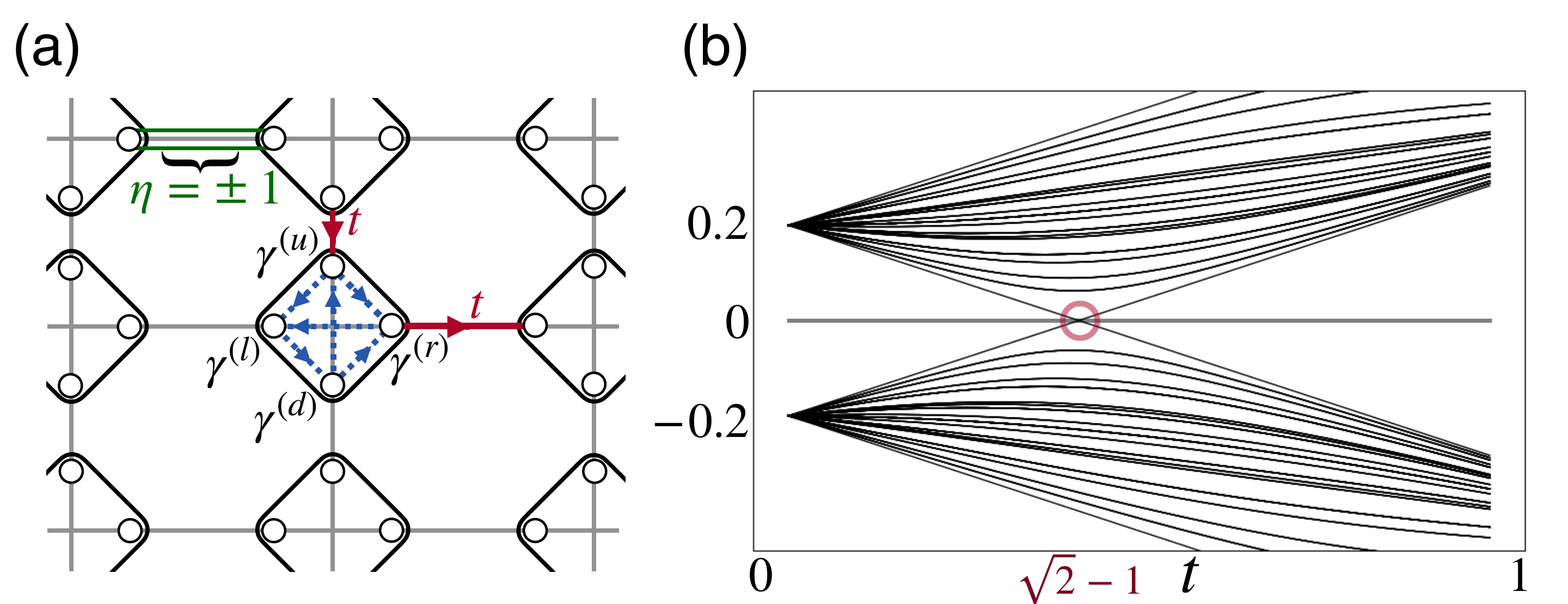}\caption{Hamiltonian $\hat{H}_{\text{MH},\ \alpha}\left[t;\ \eta\right]$ for class D
Majorana fermions in the presence of the $\mathbb{Z}_{2}$ gauge field $\eta$ (green
double line) (a), and the spectrum of $H_{\text{MH},\ \text{PP}}\left[t;\ \eta=+1\right]$
as a function of $t$ for all $\eta$ positive (b). In (a), each site
contains four Majorana modes, denoted by $\gamma^{\left(u,\ d,\ r,\ l\right)}$.
The intercell hopping constant is $t$ (red solid line), with arrows
indicating the hopping direction, e.g., $t\gamma_{\boldsymbol{r}}^{\left(u\right)}\gamma_{\boldsymbol{r}+\hat{e}_{y}}^{\left(d\right)}$.
The intra-cell interaction terms have a coupling constant $1$ (blue
dashed lines). 
(b) shows the energy spectrum of a $16\times 16$-site $H_{\text{MH},\ \text{PP}}\left[t;\ \eta=+1\right]$, with a zero mode at $t=\sqrt{2}-1$, while the spectra for $H_{\text{MH},\ \alpha}$ for $\alpha\neq \text{PP}$
displays a finite-size gap.
\label{fig:Hamiltonian_2d_pwave}}
\end{figure}

\section{Majorana representation \label{sec:TSC}}

In this section, we derive one of our main results: The error correctable phase in the toric code corresponds to the topological phase in class D disordered Majorana fermions with $\mathbb{Z}_2$ gauge fields.
This is established by showing that the CI acts as a mixed-state topological
order parameter in the fermion representation, introduced for all symmetry classes in \cite{huang2024arxiv}; see Eqs.~(\ref{eq:coherent_sc},\ref{eq:IcP}) for the key formulas. This link is reflected in excitations: For example, $e-m$ bound
states correspond to Majorana zero modes (with $e/m$ for excitations
with $A_{s}=-1$/$B_{p}=-1$), and $m$ excitations to $\mathbb{Z}_{2}$ vortices. At low error rates, $e$ and $m$ particles are sparsely distributed, allowing the underlying string to be inferred by pairing nearby particles. As the error probability increases, the density of $e$ and $m$ particles rises, and long-length string excitations become equally likely. Beyond a critical error rate, the string connecting $e$ and $m$ particles becomes ambiguous, rendering error correction impossible. In the disordered Majorana representation, this corresponds to the proliferation of $\mathbb{Z}_2$ vortices, triggering a topological phase transition when their density exceeds a critical value. Consequently, the wavefunction of Majorana zero modes trapped in vortices overlaps with those due to disorder, lifting this state to finite energy. This implies a connection between the bulk-vortex correspondence in the Majorana representation and the error-correctable code space in the toric code, which we will make precise below.

\subsection{Representing information measures in terms of Majorana fermions}
We now derive the corresponding Majorana Hamiltonian (Fig. \ref{fig:Hamiltonian_2d_pwave} (a) and Eq. \eqref{eq:TSC_ham}) and connect it to two information theoretic quantities, the R\'enyi-2 CI $I_c^{(n=2)}$ and the full CI $I_c = I_c^{(n\to 1^+)}$. To this end, we first express these quantities as an Ising model coupled to a $\mathbb{Z}_2$ gauge field $\eta$ via Eq.~\eqref{eq:Renyi_n_CI_2qubit} and Eq.~\eqref{eq:Renyi_2_Ising}, setting $\eta = +1$ homogeneously in the R\'enyi-$2$ case.  This yields (see Eq.~\eqref{eq:Renyi_2_Ising}),
\begin{equation}
I_{c}^{\left(2\right)}=2\log_{2}\frac{2Z_{\text{IM}}\left[J\right]}{\sum_{\alpha}Z_{\text{IM},\ \alpha}\left[J\right]},\ \text{with}\ \tanh J=\left(1-2p\right)^{2},\label{eq:CI_Renyi_2_IM}
\end{equation}
while in the $n\rightarrow 1^+$ limit, Eq.~\eqref{eq:Renyi_n_CI_2qubit} becomes
\begin{equation}
I_{c}=2\langle\langle\log_{2}\frac{2 Z_{\text{IM}}\left[K;\ \eta\right]}{\sum_{\alpha}Z_{\text{IM},\ \alpha}\left[K;\ \eta\right]}\rangle\rangle,\ \text{with}\ \tanh K= \left(1-2p\right),\label{eq:CI_Renyi1_IM}
\end{equation}
where the double bracket $\langle\langle \dots \rangle\rangle$ denotes averaging over $\mathbb{Z}_2$ gauge field configurations with probability distribution $P[K;\ \eta] = \prod_{\langle i,j\rangle} \frac{e^{K \eta_{ij}}}{2\cosh K}$, introduced to simplify the notation for the sum over random bond configurations.
The $\mathbb{Z}_2$ gauge field $\eta$ enters the Hamiltonian via minimal coupling 
\begin{eqnarray}
Z_{\text{IM},\ \alpha}[K;\ \eta]&\equiv \sum_{\{\sigma\}}e^{-H_{\text{IM},\ \alpha}[K;\ \eta]},
\\\nonumber
H_{\text{IM},\ \alpha}[K;\ \eta]&\equiv -K\sum_{\langle i,j \rangle }\eta_{ij}\sigma_i \sigma_j,
\end{eqnarray}
preserving the $\mathbb{Z}_2$ gauge symmetry, $\sigma_i\rightarrow \sigma_i\tau_i,\ \eta_{ij}\rightarrow \tau_i\eta_{ij}\tau_j$ for $\tau_i =\pm1$.
As above, the subscript $\alpha$ refers to periodic/anti-periodic boundary conditions (denoted by 'PP, AP, PA, AA'), with 'PP' being the default unless explicitly stated hereafter. 

The Ising model has a Majorana representation \cite{samuel1980jmp,dotsenko1983ap, cho1997prb,shalaev1994pr,read2000prbrbim,gruzberg2001prb,merz2002prb,wegner2016springer, fradkin2013cambridge, shankar2017cambridge, venn2023prl, wille2024prr, behrends2024prr}. Here we follow the method in Refs. \cite{samuel1980jmp, wegner2016springer} (see Appendix \ref{sec:fermionization} for details) to map $Z_{\text{IM}}$ to  class D Majorana fermions coupled to a $\mathbb{Z}_{2}$ gauge field. Unlike the Jordan-Wigner transformation, this fermionization method is exact for classical spin models, valid for any system size. 
The resulting Majorana Hamiltonian is (see Fig. \ref{fig:Hamiltonian_2d_pwave})
\begin{equation}
-i\hat{H}_{\text{MH}}\left[t;\ \eta\right]=\hat{A}_{\text{inter}}+\hat{A}_{\text{intra},\ 1}+\hat{A}_{\text{intra},\ 2},\label{eq:TSC_ham}
\end{equation}
which contains four Majorana modes per unit cell, labeled $\hat{\gamma}^{(u, d, l, r)}$ for the four directions (up, down, left, right, see Fig. \ref{fig:Hamiltonian_2d_pwave} (a)). $\hat{A}_{\text{inter}}$ represents inter-cell hopping (red solid line in Fig. \ref{fig:Hamiltonian_2d_pwave} (a)), 
\begin{equation}
\hat{A}_{\text{inter}}=t\left(\eta_{\boldsymbol{r},\ \boldsymbol{r}+\hat{e}_y}\hat{\gamma}_{\boldsymbol{r}}^{\left(u\right)}\hat{\gamma}_{\boldsymbol{r}+\hat{e}_{y}}^{\left(d\right)}+\eta_{ \boldsymbol{r}+\hat{e}_x,\ \boldsymbol{r}}\hat{\gamma}_{\boldsymbol{r}+\hat{e}_{x}}^{\left(l\right)}\hat{\gamma}_{\boldsymbol{r}}^{\left(r\right)}\right),
\end{equation}
and the $\alpha$ boundary condition of $\hat{\gamma}$ is implemented by changing the sign of the hopping constant across boundaries, equivalent to inserting an extra $\mathbb{Z}_2$ flux line. $\hat{A}_{\text{intra}, 1}$ and $\hat{A}_{\text{intra}, 2}$ account for intra-cell couplings for nearest and next-nearest neighbors (blue dashed lines in Fig. \ref{fig:Hamiltonian_2d_pwave} (a)), respectively,
\begin{eqnarray}\label{eq:t1t2}
\ensuremath{\hat{A}_{\text{intra},\ 1}&=&\left(\hat{\gamma}_{\boldsymbol{r}}^{\left(l\right)}\hat{\gamma}_{\boldsymbol{r}}^{\left(u\right)}+\hat{\gamma}_{\boldsymbol{r}}^{\left(d\right)}\hat{\gamma}_{\boldsymbol{r}}^{\left(l\right)}+\hat{\gamma}_{\boldsymbol{r}}^{\left(r\right)}\hat{\gamma}_{\boldsymbol{r}}^{\left(d\right)}-\hat{\gamma}_{\boldsymbol{r}}^{\left(u\right)}\hat{\gamma}_{\boldsymbol{r}}^{\left(r\right)}\right)},
\nonumber\\
\hat{A}_{\text{intra},\ 2}&=&\left(\hat{\gamma}_{\boldsymbol{r}}^{\left(u\right)}\hat{\gamma}_{\boldsymbol{r}}^{\left(d\right)}+\hat{\gamma}_{\boldsymbol{r}}^{\left(l\right)}\hat{\gamma}_{\boldsymbol{r}}^{\left(r\right)}\right).
\end{eqnarray}
For the R\'enyi-$2$ and the CI cases, the hopping constant $t$ in $\hat{H}_{\text{MH},\ \alpha}$ is, respectively,
\begin{equation}\label{eq:t_1_2_CI}
\begin{cases}
I_{c}^{\left(2\right)}: & t\rightarrow t_{2}=\left(1-2p\right)^{2}\\
I_{c}: & t\rightarrow t_1=\left(1-2p\right)
\end{cases}.
\end{equation}
Thus, $I_c^{(2)}$ (Eq.~\eqref{eq:CI_Renyi_2_IM}) corresponds to  \textit{clean} class D Majorana fermions with all $\eta$ positive, while the CI (Eq. \eqref{eq:CI_Renyi1_IM}) represents a \textit{disordered} one with random $\mathbb{Z}_2$ gauge fields. The corresponding spectrum for the clean case (Fig. \ref{fig:Hamiltonian_2d_pwave} (b)) shows an exact zero-energy point at $t_2 = \sqrt{2} - 1$ for any system size, leading to the  critical error rate of the R\'enyi-$2$ CI,
\begin{eqnarray}
    p_c = \frac{1}{2}\left(1 - \sqrt{\sqrt{2} - 1}\right).
\end{eqnarray}
This reproduces the result from the duality analysis. 

In the remainder of this section, we focus on the CI ($n\rightarrow 1^+$). In the Majorana representation, remarkably the CI is related to fermion parity (see App. \ref{sec:fermionization} for details), with the exact formula valid for any system size, 
\begin{equation}
I_{c}=2\lim_{\beta\rightarrow0^{+}}\langle\langle\log_{2}\left(1-2\frac{\mathcal{P}_{\text{PP}}\left[\beta,\ t_{1};\ \eta\right]}{\sum_{\alpha}\mathcal{P}_{\alpha}\left[\beta,\ t_{1};\ \eta\right]}\right)\rangle\rangle,\label{eq:coherent_sc}
\end{equation}
which is an even function of $\mathcal{P}_\alpha$, remaining invariant under the transformation $\mathcal{P}_\alpha\rightarrow -\mathcal{P}_\alpha$ for all $\alpha$. The high-temperature limit ($\beta \rightarrow 0^+$) reflects the classical nature of the underlying spin model. Here, $\mathcal{P}_{\alpha}\left[\beta,\ t_{1};\ \eta\right]$ represents
the expectation value of the fermion parity operator  $\left(-\right)^{\hat{Q}}$  for a class D Majorana model ($(-1)^{\hat Q} = \prod_{\boldsymbol{r}}\prod_{a=u,d,l,r}\hat{\gamma}_{\boldsymbol{r}}^{(a)}$), and
\begin{equation}\label{eq:hilbertr}
\mathcal{P}_{\alpha}\left[\beta,\ t_{1};\ \eta\right]\equiv\text{Tr}\left[e^{-\beta\hat{H}_{\text{MH},\ \alpha}\left[t_{1};\ \eta\right]}\left(-\right)^{\hat{Q}}\right] \equiv \langle\left(-\right)^{\hat{Q}}\rangle_\alpha ,\ 
\end{equation}
with $t_{1}$ given in Eq.~\eqref{eq:t_1_2_CI}. 
Importantly, the fermion parity operator changes the \textit{temporal} boundary condition of Majorana fermions from anti-periodic to periodic. Physically, it corresponds to inserting a temporal $\mathbb{Z}_2$ flux. On the other hand, the index $\alpha$ tracks different \textit{spatial} boundary conditions, indicating the presence of a spatial $\mathbb{Z}_2$ flux. The building block $\mathcal{P}_\alpha$ can thus be interpreted as follows~\cite{huang2024arxiv, huang2022prb}: The temporal flux is utilized to probe the (nonlinear) response of a system subjected to spatial fluxes, i.e., twisted boundary conditions, reflecting the presence or absence of topologically protected zero modes induced by topological defects. This is detailed in Sec.~\ref{secC} below, where we also establish a precise connection to mixed-state topological order parameters for fermions, likewise featuring the building blocks $\mathcal{P}_\alpha$.

While we relegate the derivation of Eq.~\eqref{eq:coherent_sc} to Appendix~\ref{sec:fermionization}, it is worth sketching the key steps: The Ising partition function Eq.~\eqref{eq:CI_Renyi_2_IM} is reformulated as a Majorana path integral, evaluating to a Pfaffian. This connects to  $\mathcal{P}_\alpha$ in the high temperature limit via
\begin{equation}\label{eq:pfaffian_fp}
\text{Pf}\left(-i H_{\text{MH},\ \alpha}\left[t_{1};\ \eta\right]\right) =\lim_{\beta\rightarrow0^{+}}\left(\frac{1}{\beta}\right)^{2N} \mathcal{P}_{\alpha}\left[\beta,\ t_{1};\ \eta\right],
\end{equation}
with $H_{\text{MH},\ \alpha}$ the first quantized counterpart of $\hat{H}_{\text{MH},\ \alpha}$. The prefactor $\beta^{-2N}$ balances the dimension of $\mathcal{P}_\alpha$ and the Pfaffian, and cancels out in the ratio $\frac{\mathcal{P}_{\text{PP}}}{\sum_\alpha \mathcal{P}_\alpha}$ relevant for the CI. Equations~(\ref{eq:coherent_sc},\ref{eq:hilbertr}) thus follow by recasting the Pfaffian associated with the Ising partition function in terms of a trace over the Majorana Hilbert space.  Crucially, the fermion parity operator arises naturally in this representation, reflecting the bosonic nature of the spin model. This mechanism explains the role of the fermion parity operator in the CI, and underlies the connection to the mixed-state topological order parameter discussed below. 

In the subsequent subsections, we further evaluate  Eq.~\eqref{eq:coherent_sc} in the case of a single site $N=1$, and in the thermodynamic limit $N\to\infty$. The first case will sharpen the role of boundary conditions, while in the second one, we establish an even more direct relation of the CI to the mixed-state topological order parameters of fermions. We also connect the zero crossing of the CI to the point of vanishing vortex fugacity.

\subsection{Illustration: Coherent information in the single site limit}
\begin{figure}[t!]
\includegraphics[scale=0.18]{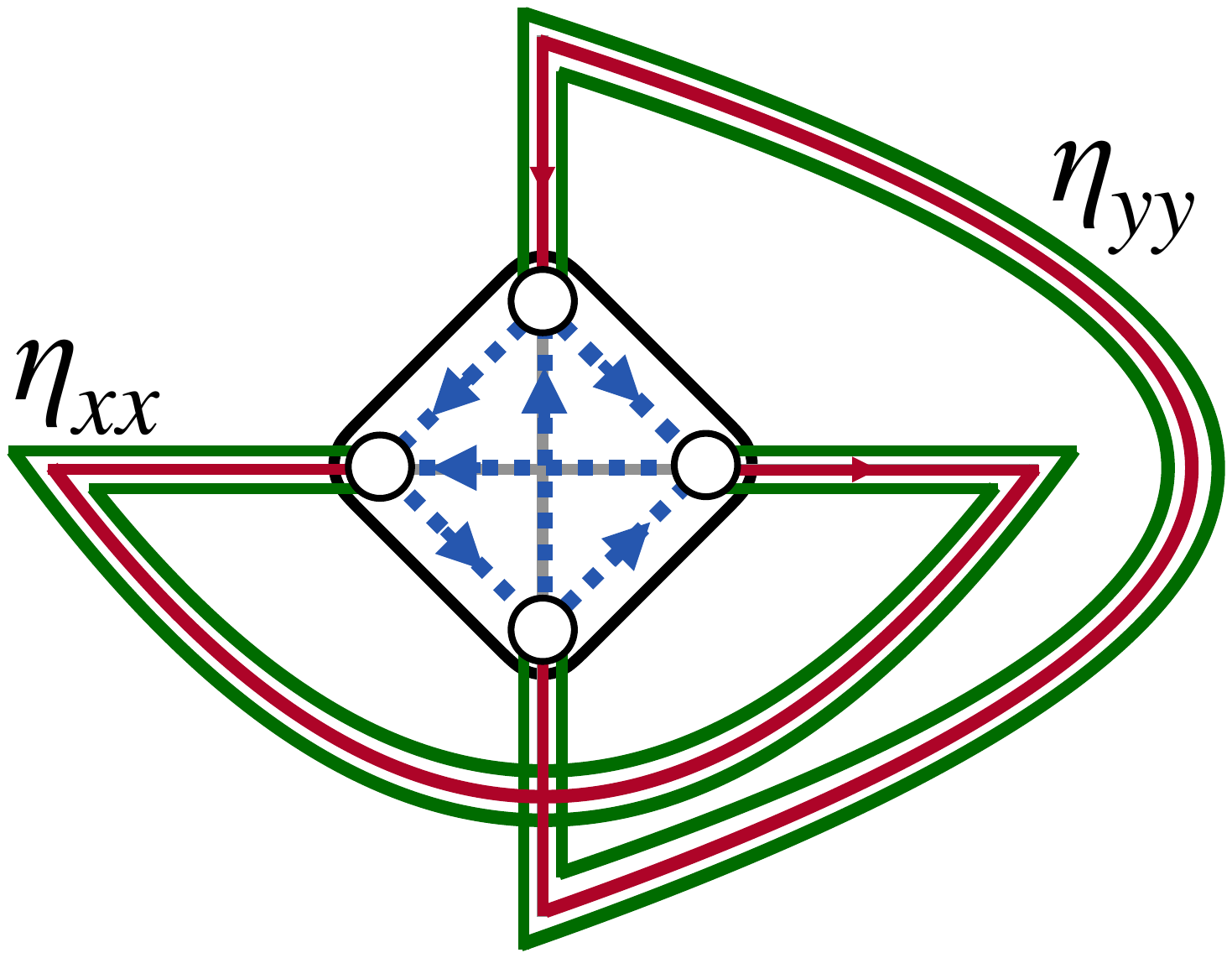}\caption{Illustration of a single-site class D superconductor setup, showing
the Hamiltonian structure with $\mathbb{Z}_2$ flux line $\eta_{xx}$ ($\eta_{yy}$) along the
$x$- ($y$-) direction. \label{fig:1_qubit}}
\end{figure}

We consider the limit of a single site, corresponding to two logical qubits. In this limit, the Hamiltonian
$H_{\text{MH},\ \alpha}$ becomes (see Fig. \ref{fig:1_qubit} for the setup),
\begin{widetext}
\begin{equation}
-iH_{\text{MH},\ \alpha}\left[t_{1};\ \eta_{xx},\ \eta_{yy}\right]=\frac{1}{2}\left(\begin{array}{cccc}
0 & t_{1}\left(-1\right)^{\alpha_{y}}\eta_{yy}+1 & -1 & -1\\
-\left[t_{1}\left(-1\right)^{\alpha_{y}}\eta_{yy}+1\right] & 0 & -1 & 1\\
1 & 1 & 0 & -\left[t_{1}\left(-1\right)^{\alpha_{x}}\eta_{xx}+1\right]\\
1 & -1 & t_{1}\left(-1\right)^{\alpha_{x}}\eta_{xx}+1 & 0
\end{array}\right),
\end{equation}
\end{widetext}
where $\eta_{xx}$ and $\eta_{yy}$ denote the $\mathbb{Z}_{2}$
flux line along $x$ and $y$ direction, respectively. $\left(\alpha_{x},\ \alpha_{y}\right)$
is for boundary condition, with $\alpha_{x/y}=0,1$ for periodic ($0$)
and anti-periodic ($1$) boundary condition along $x/y$ direction.
The fermion parity can be evaluated explicitly via Eq.~\eqref{eq:pfaffian_fp}, i.e., 
\begin{multline}
\lim_{\beta\rightarrow 0^+}\left[\left(\frac{1}{\beta}\right)^{2N}\mathcal{P}_\alpha\right] = \text{Pf}\left(-iH_{\text{MH},\ \alpha}\right)=\\ -\frac{1}{4}\left[t_{1}\left(-1\right)^{\alpha_{x}}\eta_{xx}+1\right]\left[t_{1}\left(-1\right)^{\alpha_{y}}\eta_{yy}+1\right]+\frac{1}{2},
\end{multline}
and 
\begin{equation}
\lim_{\beta\rightarrow 0^+}\left[\sum_\alpha \left(\frac{1}{\beta}\right)^{2N}\mathcal{P}_\alpha\right]=\sum_{\alpha}\text{Pf}\left(-iH_{\text{MH},\ \alpha}\right)=1,
\end{equation}
from which one can infer the CI, 
\begin{equation}\label{eq:CI_1site_flux}
I_{c}=2\langle\langle\log_{2}\left[\frac{e^{K\left(\eta_{xx}+\eta_{yy}\right)}}{\left(2\cosh K\right)^{2}}\right]\rangle\rangle+2.
\end{equation}
This result matches the CI for two qubits under both bit-flip and phase errors (Eq.
(\ref{eq:CI_1qubit})), 
\begin{equation}
I_{c}=4\left[p\log_{2}p+\left(1-p\right)\log_{2}\left(1-p\right)\right]+2,
\end{equation}
by using the identity $p=\frac{e^{-K}}{2\cosh K}$. This indicates
that in the Majorana representation, the code space consists
of the different boundary condition sectors.

This toy model provides a physical interpretation of the CI as the free energy cost of flux insertion. The Boltzmann weight for the spatial flux $\eta_{xx}$ (or $\eta_{yy}$) is $\frac{e^{K\eta_{xx}}}{2\cosh K}$ (or $\frac{e^{K\eta_{yy}}}{2\cosh K}$), resulting in an average free energy:
\begin{equation}
\langle\langle \log_2\left(\frac{e^{K\eta_{xx}}}{2\cosh K}\right) \rangle\rangle +\langle\langle \log_2\left(\frac{e^{K\eta_{yy}}}{2\cosh K}\right)\rangle\rangle.
\end{equation}
This relates to Eq.~\eqref{eq:CI_1site_flux} up to a constant, confirming the above interpretation.

\subsection{Coherent information as a mixed-state topological order parameter for class D Majorana fermions}\label{secC}
Here we consider the thermodynamic limit $N\to \infty$, and show that the CI equals a mixed-state topological order parameter for a class D Majorana model. We also connect the CI to the bulk-vortex correspondence.

The simplifications obtained as $N\to \infty$ equally rely on facts from  quantum information and topological  matter: On the one hand, as a quantum information quantity, $I_c$ is bounded by $\left|I_{c}\right|\leq2$, reflecting the maximum information content of the two logical qubits in the toric code. On the other hand, for the gapped phase in the thermodynamic limit, boundary effects become negligible compared to the bulk, and thus the absolute value of $\mathcal{P}_{\alpha}\left[\beta,\ t_{1};\ \eta\right]$ remains constant under different boundary conditions; in formulas, $\mathcal{P}_\alpha = \sign(\mathcal{P}_\alpha)\times |\mathcal{P}_{\text{PP}}|$. Hence, to ensure $|I_c|\leq 2$, all $\mathcal{P}_{\alpha\neq \text{PP}}$ are equal, allowing only $\mathcal{P}_{\text{PP}}$ or $\prod_{\alpha \neq {\text{PP}}}\mathcal{P}_\alpha$ to change sign as the error rate varies. This yields
\begin{equation}
1-2\frac{\mathcal{P}_{\text{PP}}}{\sum_\alpha \mathcal{P}_\alpha}=2^{-\text{sign}(\prod_{\alpha}\mathcal{P}_{\alpha})},
\end{equation}
which remains an even function of $\mathcal{P}_\alpha$ (i.e., unchanged under $\mathcal{P}_\alpha \rightarrow -\mathcal{P}_\alpha$ for all $\alpha$).
We thus establish a connection between $I_c$ and the mixed-state topological order parameter, defined as $\langle\langle\text{sign}\left(\frac{\mathcal{P}_{\text{PP}}\ \mathcal{P}_{\text{AA}}}{\mathcal{P}_{\text{AP}}\ \mathcal{P}_{\text{PA}}}\right)\rangle\rangle$ in Ref.~\cite{huang2024arxiv}, namely
\begin{eqnarray}\label{eq:IcP}
I_{c} &=&-2\langle\langle\text{sign}\left(\frac{\mathcal{P}_{\text{PP}}\left[\beta,\ t_{1};\ \eta \right]\mathcal{P}_{\text{AA}}\left[\beta,\ t_{1};\ \eta\right]}{\mathcal{P}_{\text{AP}}\left[\beta,\ t_{1};\ \eta\right]\mathcal{P}_{\text{PA}}\left[\beta,\ t_{1};\ \eta\right]}\right)\rangle\rangle. \nonumber \\
\end{eqnarray} 
Remarkably, the CI in the thermodynamic limit is shown to be a topological invariant itself, and thus takes quantized values. Specifically, the topological phase for mixed fermion states is characterized by a negative value of the order parameter, and thus corresponds to the error-correctable phase in the toric code, with $I_c = 2$ indicating the information content of the two logical qubits in the toric code. 

\begin{figure}[t!]
\includegraphics[scale=0.21]{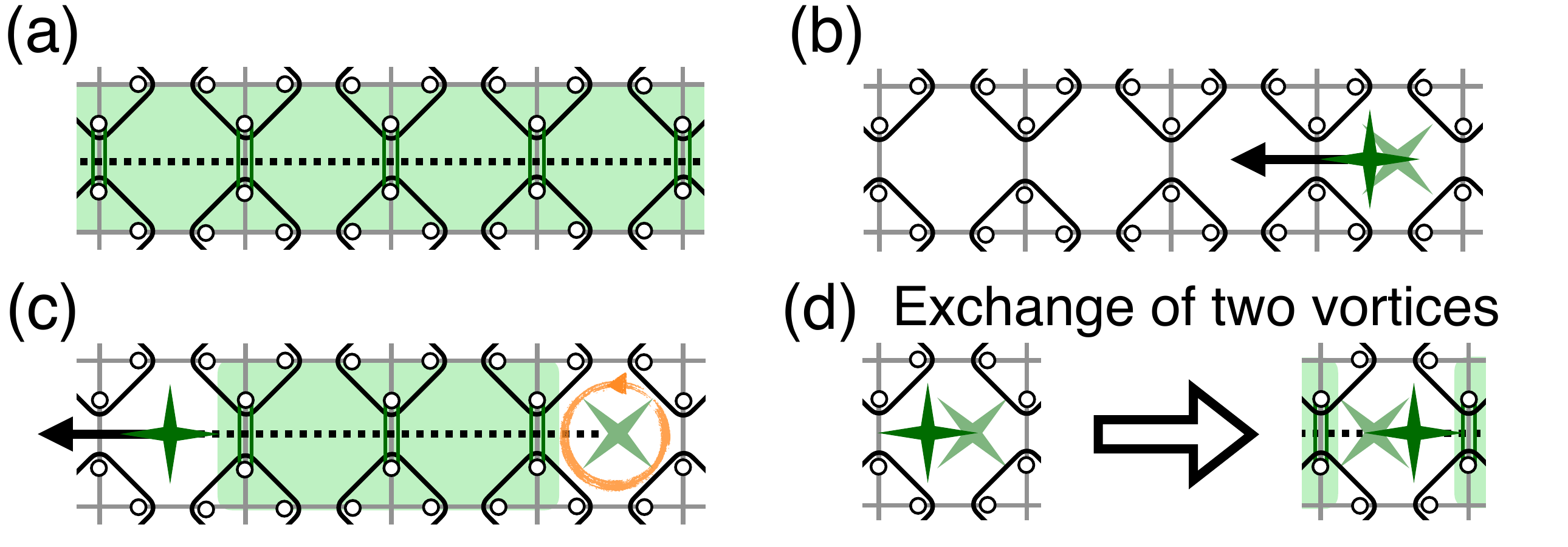}\caption{Twisted boundary condition (a) and vortex braiding (b-d): A twisted
boundary condition corresponds to the insertion of a $\mathbb{Z}_{2}$
flux line (shaded area in (a)) and can be generated by exchanging
a pair of vortices (b-d). Each four-point star represents a vortex, where the $\mathbb{Z}_2$ gauge field $\eta$ picks up a negative sign around a plaquette (orange circle in (c)). When two vortices are exchanged (d), the fermion parity operator changes sign depending on the number of Majorana zero modes (MZMs) in each vortex \cite{FN_MZM}: a negative (positive) sign appears for an odd (even) number of MZMs, since fermion parity is a product of Majorana operators. Hence, the relative sign in fermion parity, evaluated with or without twisted boundary conditions, captures the vortex braiding phase (d), reflecting the bulk-vortex correspondence. This establishes the coherent information as a topological order parameter, as it is tied to the fermion parity operator and consequently encodes the vortex braiding phase. \label{fig:braiding}}
\end{figure}

The mixed-state topological phase exhibits bulk-vortex correspondence, where the error-correctable code space in the toric code is reflected in the Majorana representation as a Majorana zero mode trapped in a vortex. This can be seen from the following argument: The relative sign of the fermion parity under different boundary conditions (entering as $\frac{\mathcal{P}_{\text{PP}}}{\sum_\alpha \mathcal{P}_\alpha}$), probes the vortex braiding phase and thus indicates whether an even or odd number of Majorana zero modes are trapped in vortices. Namely, changing the boundary condition corresponds
to inserting a $\mathbb{Z}_{2}$ flux line (see Fig. \ref{fig:braiding} (a)), equivalent to dragging one vortex around another before annihilation \cite{wen2004oxford,oshikawa2006prl} (see Fig. \ref{fig:braiding} (b-d) for an illustration). Thus, the sign of $I_c$ detects this vortex braiding phase, acting as an order parameter for the bulk-vortex correspondence. The braiding phase becomes ambiguous at $I_c=0$, marking the onset of a topological phase transition and thus demonstrating the self-duality at the critical point, completing the discussion at the beginning of Sec.~\ref{sec:selfdu}. Additional numerical results are provided in Appendix \ref{supp_sec:numerics_tpt}.

The  connection of the CI to mixed-state topology of fermions \cite{huang2024arxiv} established above allows for several further insights which should be pointed out: 
First, the associated transitions proceed without thermodynamic singularities, such as divergent length and time scales -- while for pure states, topological and thermodynamic phase transitions coincide. What these order parameters detect is indeed boundary effects, here, the loss of two Majorana zero modes. This aligns with the fact that the physics of the decoding phase transition is associated to the loss of two logical qubits stored in an extensive number $N\to\infty$ of physical qubits. Second, the  simplification of Eq.~\eqref{eq:IcP}, compared to the starting point Eq.~\eqref{eq:coherent_sc}, rests on the thermodynamic limit. In terms of a physical interpretation, it is only in this limit where the hybridization of Majorana zero modes trapped in defects vanishes. Finally, it shows that the information theoretic CI, which is highly nonlinear in the density matrix of the original stabilizer code, reduces to a mixed-state topological order parameter which is linear in the fermion density matrix. However, it involves a global probe operator, $(-)^{\hat Q}$ -- clearly, the information on the fermion parity is non-locally stored in the state.

The error threshold of the toric code attains a physical interpretation as the zero vortex fugacity point in the Majorana representation. This arises from the formal similarity between our high-temperature limit and the disordered topological insulator/superconductors in their ground state, both of which involve zero Matsubara frequency: In disordered fermions at zero temperature, this is due to the static, zero frequency limit. In the case of our finite temperature Majorana fermions, it is the fermion parity operator insertion which changes the temporal boundary conditions of the partition function to periodic ones as mentioned above, and thus allows for a zero Matsubara frequency mode to be present. 

\begin{figure}[t!]
\includegraphics[scale=0.5]{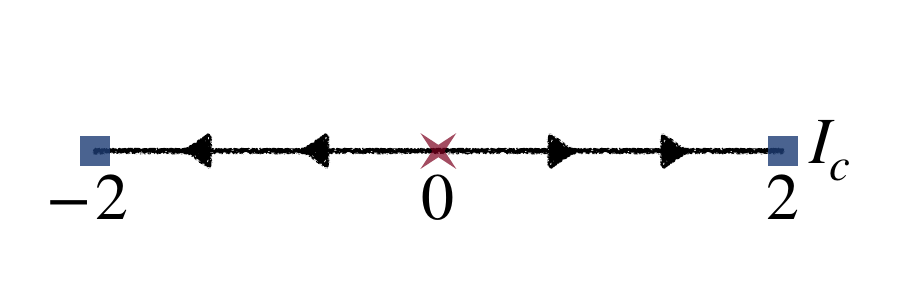}\caption{Schematic of the renormalization group flow for the coherent information $I_c$ in the toric code with both bit-flip and phase errors, assuming two phases: there are three fixed points—stable at $I_c = \pm 2$ in the gapped phase, and unstable at $I_c = 0$ at the critical point with zero vortex fugacity. As the system size increases, $I_c$ flows toward the stable points $I_c = \pm 2$, while $I_c = 0$ remains unchanged.
\label{fig:RG_flow}}
\end{figure}

In the disordered scenario, the critical point is identified by zero vortex fugacity (denoted by $u$) \cite{fu2012prl, konig2012prb, altland2014prl, altland2015prb}. Themodulus of $u$ represents the statistical weight for vortex creation, while its sign is an indicator of topology \cite{fu2012prl, konig2012prb, altland2014prl, altland2015prb}: Negative $u$ signals a topological phase with non-trivial bulk-vortex correspondence, and positive $u$  a normal phase. Thus, we observe that  the zero vortex fugacity point coincides with the zero crossing of the CI, $u \sim I_c$. Quantitatively, the critical point of disordered fermions is captured by the following RG equation for the fugacity \cite{read2000prb,fu2012prl,konig2012prb, altland2014prl, altland2015prb},
\begin{equation}
\frac{du}{d\ln L}\propto u,
\end{equation}
where $L$ is the  RG length scale. This shows that the zero-fugacity point ($u=0$), and hence zero CI, is located exactly at the RG fixed point, remaining unrenormalized as the system size increases (see Fig.~\ref{fig:RG_flow} for the schematic RG flow for $I_c$). This fits  numerical results showing tiny finite-size drift at the zero-CI point \cite{colmenarez2024prr} (see also Appendix \ref{supp_sec:numerics_tpt} for additional numerical results). It rationalizes that indeed the CI can effectively probe the error threshold in small systems.

\section{Numerical results for other models \label{sec:numerical_result}}
\begin{figure}[t]
\includegraphics[scale=0.18]{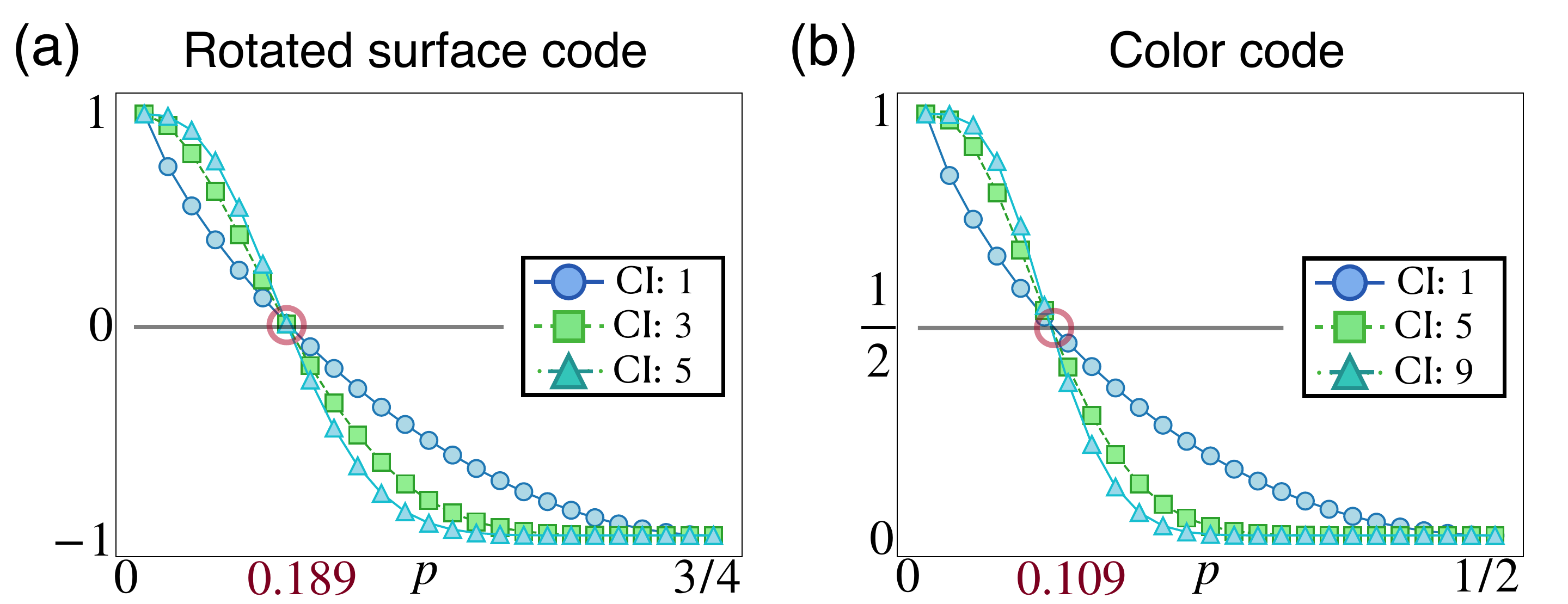}

\caption{Numerical results for the coherent information (CI) in (a) the rotated surface code under depolarizing noise and (b) the color code under bit-flip errors, evaluated across different code distances ($1$, $3$, and $5$ for the surface code; $1$, $5$, and $9$ for the color code). The CI is computed by diagonalizing the density matrix \cite{colmenarez2024prr}. The error threshold, determined from the crossing point of the CI, is $0.189$ for the rotated surface code and $0.109$ for the color code, aligning with previously reported results \cite{katzgraber2009prl, bombin2012prx, terhal2015rmp}. \label{fig:color_planar}}
\end{figure}

Through the lens of the Majorana representation, we identify vortex proliferation as the mechanism driving the decoding phase transition. The connection of the threshold to a renormalization group fixed point suggests that the CI is an efficient tool for determining the latter. Although our analysis focuses on the toric code model under bit-flip and phase errors, we expect this mechanism to be general, as all 2D topological stabilizer codes are equivalent to multiple copies of the toric code \cite{yoshida2011aop,bombin2012njp, kubica2015njp}. We support this by numerical results, i.e., applying CI across various code models with small code distances (see Fig. \ref{fig:color_planar}), and accurately reproducing known error thresholds. This includes: (a) The rotated surface code under depolarizing noises; (b) The triangular 4.8.8. color code under bit-flip errors (see e.g., Ref. \cite{pedro2022pra,colmenarez2024prr} for details on these models). %The CI is computed numerically by diagonalizing the mixed-state density matrix as done in Ref.~\cite{colmenarez2024prr}.

\section{Conclusion and outlook \label{sec:conclude_outlook}}
In this work, we have constructed a connection of a quantum information theoretic quantity, the coherent information (CI), and the physics of topological superconductors in mixed quantum states. This connection is enabled by an exact mapping of the CI itself to Majorana fermions, valid for any system size and carefully tracking the boundary conditions. In the thermodynamic limit, the CI reduces exactly to a mixed-state topological order parameter. Remarkably, the latter is linear in the fermion density matrix, and it takes quantized values reflecting the sealing of quantum information below the error threshold. One benefit of the new mapping is a rather direct interpretation of information theoretic properties - like the error correctable phase, the error threshold, and the stored information - in terms of basic many-body properties of fermions - the topologically non-trivial phase, the mixed-state phase transition, and the existence of topologically protected zero modes. 
In particular, as a consequence of the aforementioned connections, two key insights regarding the toric code decoding phase transition follow: (i) emergent self-duality at the critical point on the Nishimori line of the random-bond Ising model, and (ii) tiny finite-size effects in the CI near the critical point, stemming from the connection of the CI zero crossing to a renormalization group fixed point.
%In this work, we have constructed a connection of a quantum information theoretic quantity, the CI, and the physics of topological superconductors in mixed quantum states.  One benefit of the new mapping is a rather direct interpretation of information theoretic properties - like the error correctable phase, the error threshold, and the stored information - in terms of basic many-body properties of fermions - the topologically non-trivial phase, the mixed-state phase transition, and the existence of topologically protected zero modes. In particular, the basic building block in the fermion language - the mixed-state topological order parameter - is linear in the density matrix.

 Our work has focused on the most paradigmatic model, the toric code under bit-flip and phase errors. An important direction of future research concerns charting the generality of the results obtained here. As mentioned, it can be expected that the CI remains an efficient tool for other two-dimensional topological stabilizer codes, as they are equivalent to multiple copies of the toric code \cite{yoshida2011aop,bombin2012njp, kubica2015njp}; we have provided numerical support for this conjecture. More broadly, mappings of complex quantum spin models to fermions are ubiquitous, when it comes to the bulk properties of the systems. To name only a few, the quantum Ising model \cite{sachdev2011cambridge, fradkin2013cambridge, shankar2017cambridge} (or repetition code, in quantum information language), the cluster state \cite{briegel2001prl,verresen2017prb}, and the XZZX code \cite{ataides2021nc, klocke2022prb, hauser2024arxiv}, all with equivalent Majorana representations; these mappings can be formulated in great generality using tensor network techniques~\cite{wille2024prr,wille2024arxiv}. Recently, motivated by the prospects of state-of-the-art quantum devices, there is a surge of research activity, studying such quantum spin models using quantum information measures like the CI. This becomes necessary, for example, in setups where quantum measurements compete with decoherence processes \cite{fisher2023ar}, or in detecting the strong-to-weak spontaneous symmetry breaking \cite{lessa2024arxiv}; more broadly, whenever one needs to consider quantities that are non-linear in the spin model density matrix. Recasting the information theoretic quantities in terms of fermions as exemplified here, and identifying the relevant mixed-state topological order parameters \cite{huang2024arxiv}, would not only be conceptually rewarding and physically insightful. It might further give rise to practical advantages for high precision estimates of error thresholds or phase diagrams of such problems. For example, computing CI via the Majorana representation, if it exists as in the case studied here, can be achieved numerically efficiently.
Conversely, an interesting direction of research is to connect known instances of mixed-state topological phase transitions in fermion systems, including in the interacting case \cite{huang2024arxivIITPT}, to quantum information measures. Ultimately, this connection could help identifying new candidate systems for robust quantum information storage and processing.

Finally, on the spin model side of the triptych in Fig.~\ref{fig:concepts}, we have found an exact relation between the CI zero point and the self-dual point of the random-bond Ising model. Combined with the previously known equivalence between the critical point and the CI zero point in the thermodynamic limit, this establishes an emergent self-duality at the critical point. It will be intriguing to explore whether the link of vanishing CI and self-duality extends to other models, such as the random three-body Ising model associated with the 2D topological color code \cite{bombin2008pra,katzgraber2009prl}, in this way leveraging quantum information tools for new insights in statistical mechanics.

\begin{acknowledgments}
We thank Alex Altland, Jan Behrends, Benjamin B\'eri, Hidetoshi Nishimori, Simon Trebst and Guo-Yi Zhu for discussions.
Z.-M. H., M. M. and S. D. are supported by the  Deutsche Forschungsgemeinschaft (DFG, German Research Foundation) under Germany’s Excellence Strategy Cluster of Excellence Matter and Light for Quantum Computing (ML4Q) EXC 2004/1 390534769, with additional funding for Z.-M.H. and S.D. from the DFG Collaborative Research Center (CRC) 183 Project No. 277101999 - project B02.
L.C. and M.M. acknowledge funding by the U.S. ARO Grant No. W911NF-21-1-0007. 
M.M. also acknowledges funding from the European Union’s Horizon Europe research and innovation programme under grant agreement No 101114305 (“MILLENION-SGA1” EU Project), and this research is also part of the Munich Quantum Valley (K-8), which is supported by the Bavarian state government with funds from the Hightech Agenda Bayern Plus. M.M. acknowledges funding from
the ERC Starting Grant QNets through Grant No.~804247, and from the European Union’s Horizon Europe research and innovation program under Grant Agreement No. 101046968 (BRISQ). 
M.M. also acknowledges support for the research that was sponsored by IARPA and the Army Research Office, under the Entangled Logical Qubits program, and was accomplished under Cooperative Agreement Number W911NF-23-2-0216. The views and conclusions contained in this document are those of the authors and should not be interpreted as representing the official policies, either expressed or implied, of IARPA, the Army Research Office, or the U.S. Government. The U.S. Government is authorized to reproduce and distribute reprints for Government purposes notwithstanding any copyright notation herein. 
\end{acknowledgments}

\appendix

\section{Coherent information and the Knill-Laflamme condition \label{supp_sec:EC_condition_CI}}

In this section, we connect two error correction conditions: the coherent information (CI) approach by Schumacher, Nielsen and Lloyd \cite{schumacher1996pra,lyoyd1997pra}, and the Knill-Laflamme condition. While this relationship has been previously established \cite{schumacher1996pra,lyoyd1997pra, barnum1998pra}, our goal is to provide a more straightforward derivation, specifically of the coherent information condition from the Knill-Laflamme condition.

Quantum memory ($Q$) is inherently fragile due to pervasive noise, leading to potential information leakage into the environment ($E$). To make quantum devices practical, error correction is essential, but it is only feasible if no information is leaked, as captured by the error correction conditions. The first condition, derived from CI, states that error correction is possible when the residual information in $Q$ (measured by CI $I_c$) equals the initially stored information $S_Q$:
\begin{equation}
I_{c}=S_{Q}.
\end{equation}
The second condition, known as the Knill-Laflamme condition, evaluates error correctability by analyzing the code space through:
\begin{equation}
P_{c}K_{\alpha_{1}}^{\dagger}K_{\alpha_{2}}P_{c}=A_{\alpha_{1}\alpha_{2}}P_{c},
\end{equation}
where $P_{c}$ is the projector onto the code space, $K_{\alpha}$ are the Kraus operators for the noise process $\alpha$, and $A_{\alpha_{1}\alpha_{2}}$ is a real number indexed by $\alpha_{1}$ and $\alpha_2$. This condition thus states that error correction is possible if the error channel does not affect the code space. 

While these two conditions approach error correction from different perspectives, we will now quantitatively demonstrate their equivalence.

\subsection{Knill-Laflamme condition $\protect\implies$ coherent information
condition }

We show that the Knill-Laflamme condition implies the CI condition via the factorization,
\begin{equation}
S_{Q^{\prime}}=S_{Q}+S_{E^{\prime}},\label{supp_eq:S_Qp-1}
\end{equation}
where $S_{Q}$ and $S_{Q^{\prime}}$ are the von Neumann entropies of the quantum memory before and after decoherence, respectively, and $S_{E^{\prime}}$ is the von Neumann entropy of the environment. This implies that the initial quantum memory information remains intact, as shown by,
\begin{equation}
S_{Q^{\prime}}-S_{E^{\prime}}=S_{Q}\implies I_{c}=S_{Q},
\end{equation}
with $I_{c}\equiv S_{Q^{\prime}}-S_{E^{\prime}}$,
thereby reproducing the CI error correction condition.

Equation~\eqref{supp_eq:S_Qp-1} is derived by expressing $S_{Q^{\prime}}$ as:
\begin{equation}
S_{Q^{\prime}}=-\text{tr}_{\alpha}A\log_{2}A+S_{Q},\label{supp_eq:S_rhoQp-1}
\end{equation}
where $\text{tr}_{\alpha}$ traces over the subscripts $\alpha$ of $A_{\alpha_{1}\alpha_{2}}$. 
$-\text{tr}_{\alpha}A\log_{2}A$ equals $S_{E^{\prime}}$ (see \ref{suppsec:supp_Eq_S_E}
for derivations), characterizing the leakage information to the environment.
Specifically, Eq. \eqref{supp_eq:S_rhoQp-1} follows from the identities:
\begin{equation}
\rho_{Q^{\prime}}=\sum_{\alpha}K_{\alpha}P_{c}\rho_{Q}P_{c}K_{\alpha}^{\dagger},
\end{equation}
and 
\begin{equation}
S_{Q^{\prime}}=-\lim_{n\rightarrow1^{+}}\frac{1}{n-1}\text{Tr}\left(\rho_{Q^{\prime}}^{n}-1\right),
\end{equation}
where $\rho_{Q}=P_{c}\rho_{Q}P_{c}$ as $\rho_{Q}$ resides
within the code space. We thus reproduce Eq.~\eqref{supp_eq:S_rhoQp-1}, 

\begin{eqnarray}
S_{Q^{\prime}} & = & -\text{Tr}\left(\rho_{Q^{\prime}}\log_{2}\rho_{Q^{\prime}}\right)\nonumber \\
 & = & -\lim_{n\rightarrow1}\frac{1}{n-1}\left\{ \text{tr}_{\alpha}\left(A^{n}\right)\times\text{Tr}\left[\left(P_{c}\rho_{Q}P_{c}\right)^{n}\right]-1\right\} \nonumber \\
 & = & -\text{tr}_{\alpha}\left(A\log_{2}A\right)+S_{Q}.
\end{eqnarray}

\subsection{Coherent information condition $\protect\implies$ Knill-Laflamme
condition }

Derivations along this line are more technical, and offer fewer physical
insights. Therefore, we direct interested readers to Refs. \cite{schumacher1996pra, barnum1998pra}. 

\subsection{Derivation of $S_{E^{\prime}}=-\text{tr}_{\alpha}A\log_{2}A$ \label{suppsec:supp_Eq_S_E}}

We prove the identity $S_{E^{\prime}} = -\text{tr}_{\alpha}A\log_{2}A$ by using the following identity (see derivations below)
\begin{equation}
\rho_{E^{\prime}}=\sum_{\alpha_{1},\ \alpha_{2}}W_{\alpha_{1}\alpha_{2}}|\alpha_{1}\rangle\langle\alpha_{2}|,\label{supp_eq:rho_E}
\end{equation}
where $W_{\alpha_{1}\alpha_{2}}\equiv\text{Tr}\left(K_{\alpha_{1}}P_{c}\rho_{Q}P_{c}K_{\alpha_{2}}^{\dagger}\right)$
relates to $A_{\alpha_{1}\alpha_{2}}$ by matrix transpose, $W_{\alpha_{1}\alpha_{2}}=A_{\alpha_{2}\alpha_{1}}$,
and $|\alpha\rangle$'s form an orthonormal basis. From this, we confirm:
\begin{equation}
S_{E^{\prime}}=-\text{tr}_{\alpha}W\log_{2}W=-\text{tr}_{\alpha}A\log_{2}A.
\end{equation}

We derive Eq. \eqref{supp_eq:rho_E} using a purification
approach. We begin with a purified wavefunction for the composite system,
\begin{equation}
|\Psi_{RQE}\rangle=\sum_{\alpha}|\Psi_{RQ}\rangle\otimes|0_{E}\rangle,
\end{equation}
where $R$ is the reference system. This reproduces the density
matrices $\rho_{Q}$ and $\rho_{E}$, 
\begin{equation}
\begin{cases}
\rho_{Q}\equiv\text{Tr}_{R,\ E}|\Psi_{RQE}\rangle\langle\Psi_{RQE}|\\
\rho_{E}\equiv\text{Tr}_{R,\ Q}|\Psi_{RQE}\rangle\langle\Psi_{RQE}|
\end{cases}.
\end{equation}
Decoherence, described by the Kraus operators $K_{\alpha}$, transforms the state as, 
\begin{equation}
\begin{cases}
|\Psi_{RQ^{\prime}E^{\prime}}\rangle=\mathbb{I}_{R}\otimes \tilde{U}_{QE}\sum_{\alpha}|\Psi_{RQ}\rangle\otimes|0_{E}\rangle\\
\tilde{U}_{QE}\equiv\sum_{\alpha}K_{\alpha}\otimes|\alpha\rangle\langle0_{E}|
\end{cases},
\end{equation}
where $\left\{ |\alpha\rangle\right\} $ is an orthonormal basis, consistent with $\sum_{\alpha}K_{\alpha}^{\dagger}K_{\alpha}=\mathbb{I}$.
The superscript '$\prime$' denotes the decohered system. This setup produces  the correct expression for $\rho_{Q^\prime}$,
\begin{equation}
\rho_{Q^{\prime}}\equiv\text{Tr}_{R,\ E}|\Psi_{RQ^{\prime}E^{\prime}}\rangle\langle\Psi_{RQ^{\prime}E^{\prime}}|=\sum_{\alpha}K_{\alpha}\rho_{Q}K_{\alpha}^{\dagger},
\end{equation}
and confirms Eq. (\ref{supp_eq:rho_E}), 
\begin{equation}
\rho_{E^{\prime}}=\sum_{\alpha_{1},\ \alpha_{2}}W_{\alpha_{1}\alpha_{2}}|\alpha_{1}\rangle\langle\alpha_{2}|,
\end{equation}
with $W_{\alpha_{1}\alpha_{2}}\equiv\text{Tr}\left(K_{\alpha_{1}}P_{c}\rho_{Q}P_{c}K_{\alpha_{2}}^{\dagger}\right)$.

\section{Duality relation for the Ising model on a torus \label{sup_sec_Dirac_delta}}

\begin{figure}
\includegraphics[scale=0.18]{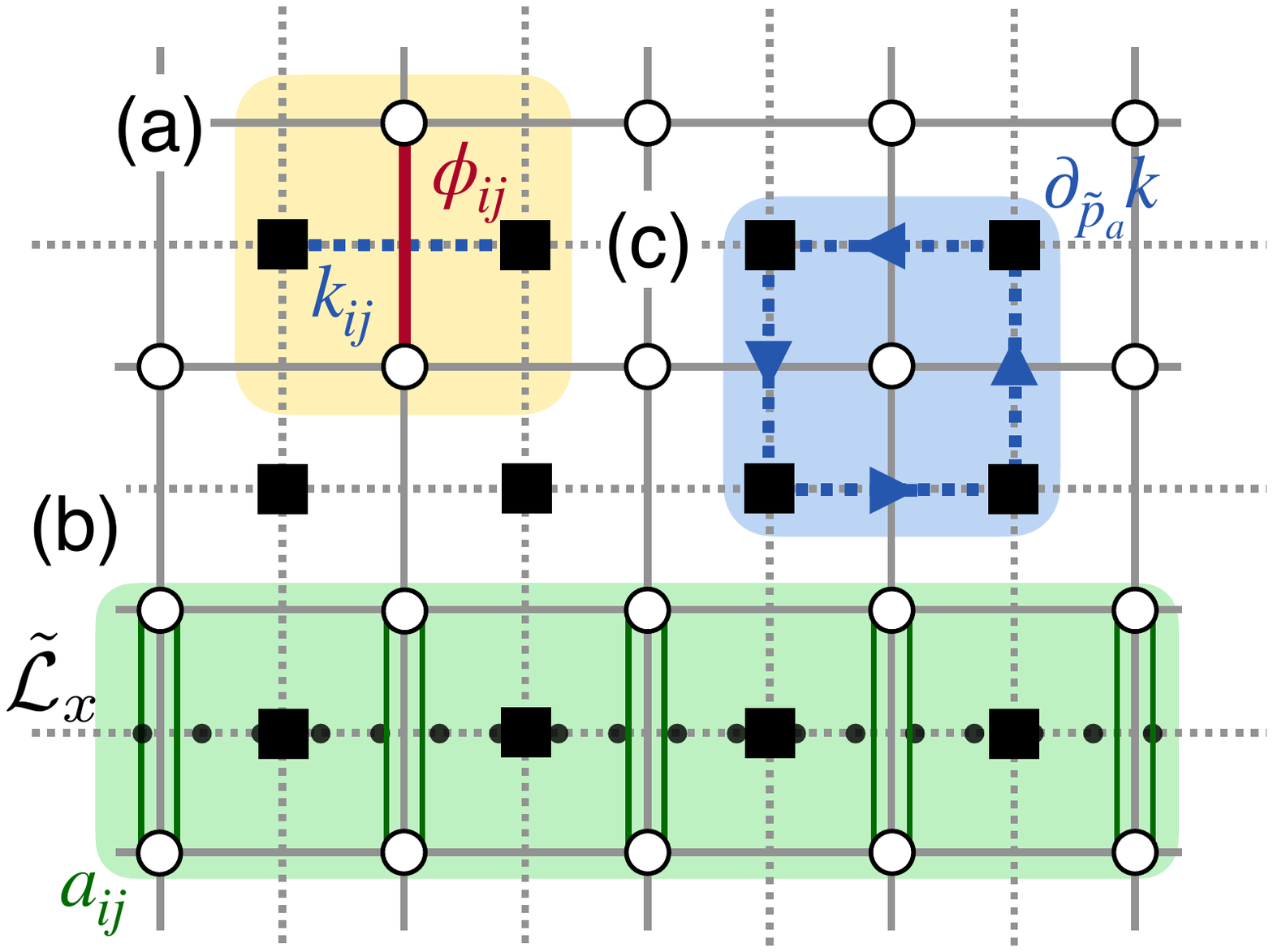}\caption{Illustration of notations: $\phi_{ij}$ and $k_{ij}$  (a), non-contractible loop $\tilde{\mathcal{L}}_x$ (b), and curl $\partial_{\tilde{p}_a}k$ (c). The solid (dashed) lines represent the (dual) lattice, with Ising spins on vertices as hollow circles (solid squares). (a) shows  variables $\phi_{ij}$ ($k_{ij}$) on the (dual) lattice. (b) illustrates a non-contractible loop $\tilde{\mathcal{L}}_x$ on the dual lattice, with the double green line representing a $\mathbb{Z}_2$ gauge field ($a_{ij}$, or equivalently $\eta_{ij}=e^{i\pi a_{ij}}$). (c) defines the curl of $k_{ij}$, where arrows indicate the sign of $k_{ij}$ (positive for right/up, negative otherwise). 
\label{supp_fig:illustration_notation}}
\end{figure}

In this section, we provide a detailed derivation of Eq. \eqref{eq:Dirac_delta_derivation}, from which we further derive the duality relation on a torus. For convenience, Eq. \eqref{eq:Dirac_delta_derivation} is restated here:
\begin{equation}\label{supp_eq:Dirac_delta_derivation}
2\sum_{\{ k \}, \ \partial k = 0} \prod_{\langle i, j \rangle} e^{i\pi k_{ij} \phi_{ij}} = \frac{1}{2}\prod_{p_a} \left[ 2 \delta \left( \partial_{p_a} \phi \ \text{mod} \ 2 \right) \right],
\end{equation}
where $\phi_{ij}$ and $k_{ij}$ are  variables on the lattice and dual lattice, respectively (red solid and blue dashed lines in Fig. \ref{supp_fig:illustration_notation} (a)). The variable $k_{ij}$ is constrained by the condition $\partial k = 0$, meaning its curl vanishes along any closed loop on the torus, including both contractible and non-contractible loops (e.g., $\tilde{\mathcal{L}}_{x/y}$ along the $x/y$ directions, black dotted lines in Fig. \ref{supp_fig:illustration_notation} (b)). Here, 'mod 2' is implied for the curl and will be omitted henceforth. 

We derive Eq.~\eqref{supp_eq:Dirac_delta_derivation} by first representing the constraints on $k_{ij}$ as Dirac-delta functions. For constraints from contractible loops, they consist of local ones associated with dual plaquettes ($\tilde{p}_a$, Fig. \ref{supp_fig:illustration_notation} (c)):

\begin{equation}
\partial_{\tilde{p}_a} k = 0 \implies  \delta\left( \partial_{\tilde{p}_a} k \right) = \frac{1}{2}\sum_{\varphi_{\tilde{p}_a} = 0, 1} e^{i \pi \varphi_{\tilde{p}_a} \left( \partial_{\tilde{p}_a} k \right)},
\end{equation}
which introduces the dual plaquette variables $\varphi_{\tilde{p}_a}$. There exist $N-1$ independent constraints on a torus, so we include only $N-1$ Dirac-delta functions, excluding the dual plaquette $\tilde{p}_N$.
For non-contractible loops, additional constraints are needed since they cannot be deformed into local one residing on dual plaquettes. We thus introduce two extra variables, $b_{d}=0,\ 1$ with $d=x,\ y$, 
\begin{equation}
\partial_{\tilde{\mathcal{L}}_{d}} k = 0 \implies \delta\left( \partial_{\tilde{\mathcal{L}}_{d}} k \right) = \frac{1}{2}\sum_{b_{d} =0, 1} e^{i \pi b_{d} \left( \partial_{\tilde{\mathcal{L}}_{d}} k \right)},
\end{equation}
where $\tilde{\mathcal{L}}_x$ and $\tilde{\mathcal{L}}_y$ are non-contractible loops along the $x$ and $y$ directions on the torus (Fig. \ref{supp_fig:illustration_notation} (b)). The auxiliary fields $b_{d} = 0,\ 1$ couple to all $k_{ij}$ along $\tilde{\mathcal{L}}_{d}$, leading to different boundary conditions for $\phi_{ij}$ along $d=x,\ y$, as shown below.

We now integrate out the unconstrained $k_{ij}$, using the identity:

\begin{equation}
\sum_{\tilde{p}_a} \varphi_{\tilde{p}_a} \left(\partial_{\tilde{p}_a} k\right) = -\sum_{\langle i, j \rangle} k_{ij}  \left( \Delta_{ij} \varphi \right),\label{supp_eq:InteBp}
\end{equation}
which is the integration by parts on a lattice, with $\Delta_{ij}$ representing the gradient along the $\phi_{ij}$ direction. Via variables $\varphi_{\tilde{p}_a}$ and $b_{x/y}$, we replace the constrained $k_{ij}$ with unconstrained ones,
\begin{eqnarray}
&&2\sum_{\{ k \}, \ \partial k = 0} \prod_{\langle i, j \rangle} e^{i \pi k_{ij} \phi_{ij}} \nonumber\\
&=& 2\sum_{\{ k \}} \prod_{\langle i, j \rangle} e^{i \pi k_{ij} \phi_{ij}} \times \left[ {\prod}_{\tilde{p}_a}^{\prime} \delta \left( \partial_{\tilde{p}_a} k \right) \right] \times \left[ \prod_{d=x,\ y} \delta \left( \partial_{\tilde{\mathcal{L}}_d} k \right) \right] \nonumber \\
&=& 2{\sum}_{\{ \varphi,\ k,\ b_d \}}^\prime e^{i \pi \sum_{\langle i, j \rangle} k_{ij} \phi_{ij}} 
\times  {\prod}^\prime_{\tilde{p}_a}\frac{1}{2}e^{i \pi  \varphi_{\tilde{p}_a} \times\left(\partial_{\tilde{p}_a} k\right)} \nonumber \\
&& \times  \prod_{d}\frac{1}{2}e^{i \pi b_d \times  \left(\partial_{\tilde{\mathcal{L}}_d} k\right)},
\label{supp_eq:Dirac_delta_varphi}
\end{eqnarray}
where the prime in ${\prod}^\prime$ and ${\sum}^\prime$ indicates exclusion of the plaquette $\tilde{p}_N$. To treat all plaquettes equally, we insert the resolution of identity:
\begin{equation}
1 = \sum_{\xi = 0, 1} \delta( \partial_{\tilde{p}_N} k - \xi ) = \sum_{\xi, \varphi_{\tilde{p}_N} = 0, 1} \frac{1}{2} e^{i \pi \varphi_{\tilde{p}_N} ( \partial_{\tilde{p}_N} k - \xi )},
\end{equation}
which simplifies Eq. \eqref{supp_eq:Dirac_delta_varphi} to:
\begin{eqnarray}
&&2\sum_{\{ k,\ \varphi,\ b_d \}} e^{i \pi k_{ij} \phi_{ij}} \times 
\left(\sum_{\xi} e^{-i \pi \varphi_{\tilde{p}_N} \xi} \right) \nonumber \\
&& \times \left(\prod_{\tilde{p}_a} \frac{1}{2}  e^{i \pi \varphi_{\tilde{p}_a}\times (\partial_{\tilde{p}_a} k)}\right)\times \left[\prod_{d=x,y} \frac{1}{2}e^{i \pi b_d \times \left(\partial_{\tilde{\mathcal{L}}_d} k\right)} \right]\nonumber \\
&=&2^{N-1} \sum_{\{ \varphi,\ b_d \}} \left[\prod_{\langle i, j \rangle} \delta \left( \phi_{ij} - \Delta_{ij} \varphi + a_{ij} \right)\right] \times 2\delta \left( \varphi_{\tilde{p}_N} \right).\nonumber\\
\end{eqnarray}
In the second line, we have integrated over $k_{ij}$ using Eq. \eqref{supp_eq:InteBp}, yielding a product of Dirac-delta functions $\delta(\phi_{ij} - \Delta_{ij} \varphi + a_{ij})$, with a prefactor $2^{N-1} = 2^{2N+1 - N - 2}$. Here, $2N$ represents the number of bonds ($k_{ij}$), and $N$ the number of dual plaquettes ($\tilde{p}_a$). The additional Dirac-delta function, $2\delta(\varphi_{\tilde{p}_N})$, results from summing over $\xi$. The variable $a_{ij} = 0, 1$ is a $\mathbb{Z}_2$ gauge field related to $\eta_{ij}$ via $\eta_{ij} = e^{i\pi a_{ij}}$, and depends on $b_d$ ($d = x, y$): $a_{ij} = 0$ for all bonds except those crossing $\tilde{\mathcal{L}}_d$, where $a_{ij} = b_d$ (green double line in Fig. \ref{supp_fig:illustration_notation} (b)). In turn, this indicates that $b_d$ represents the Wilson loop associated with $a_{ij}$.
To simplify, we use the shift symmetry of $\varphi_{\tilde{p}_a}$ and replace $2\delta(\varphi_{\tilde{p}_N})$ with an average over different $\varphi_{\tilde{p}_N}$ values, which reduces Eq. \eqref{supp_eq:Dirac_delta_varphi} to,
\begin{multline}  
    2\sum_{\{ k \}, \ \partial k = 0} \prod_{\langle i, j \rangle} e^{i\pi k_{ij} \phi_{ij}} \\
    = 2^{N-1}\sum_{\{\varphi,\ b_d\}} \prod_{\langle i, j \rangle} \delta \left( \phi_{ij} - \Delta_{ij} \varphi + a_{ij} \right).\label{supp_eq:varphi_Dirac_delta}
\end{multline}
These Dirac-delta functions indicate that $\phi_{ij}$ is a locally exact form in the bulk (away from $\tilde{\mathcal{L}}_x$), where $\phi_{ij} = \Delta_{ij} \varphi$, but with non-trivial holonomy due to $a_{ij}$. This imposes a curl-free condition on $\phi$ for contractible loops ($\partial_{p_a}\phi$), so by integrating out $\varphi_{\tilde{p}_a}$, we recover Eq. \eqref{supp_eq:Dirac_delta_derivation}.

Building on these results, one can infer the duality relation
\begin{equation}\label{supp_eq:duality_relation}
2 \tilde{Z}_{\text{IM}}[J] = \sum_{\alpha} Z_{\text{IM},\ \alpha}[J],
\end{equation}
by taking $\varphi_{\tilde{p}_a}$ as the Ising spin, with boundary conditions set by the Wilson loop $b_d$, i.e., $\cos(\pi \varphi_{\tilde{p}_a}) = \sigma_i$, where the dual plaquette $\tilde{p}_a$ corresponds to a lattice site $i$. This relation is derived by integrating out $\phi_{ij}$ in $\tilde{Z}_{\text{IM}}[J]$ using Eq. \eqref{supp_eq:varphi_Dirac_delta}. Specifically, for $\tilde{Z}_{\text{IM}}[J]$:
\begin{equation}
\tilde{Z}_{\text{IM}}[J] = \sum_{\{\phi\}} \left\{ 2 \sum_{\{k\}} \prod_{\langle i, j \rangle} \left[ \frac{1}{\sqrt{2}} e^{i\pi k_{ij} \phi_{ij}} x(\phi_{ij}) \right] \right\},
\end{equation}
we integrate out both $k_{ij}$ (using Eq. \eqref{supp_eq:varphi_Dirac_delta}) and $\phi_{ij}$, resulting in,
\begin{eqnarray}
\tilde{Z}_{\text{IM}}[J] &=& \frac{1}{2} \sum_{b_d} \left\{ \sum_{\{\varphi\}} \prod_{\langle i, j \rangle} x_{ij, \alpha}[\Delta \varphi-a_{ij}] \right\}\nonumber\\
{}&=& \frac{1}{2} \sum_{\alpha} Z_{\text{IM}, \alpha}[J],
\end{eqnarray}
where the prefactor $\frac{1}{2} = \frac{2^{N-1}}{\sqrt{2}^{2N}}$, and the boundary conditions are determined by $b_d$. This establishes the duality relation Eq. \eqref{supp_eq:duality_relation}.

\section{Fermionization via loop expansion \label{sec:fermionization}}

\begin{figure}
\includegraphics[scale=0.18]{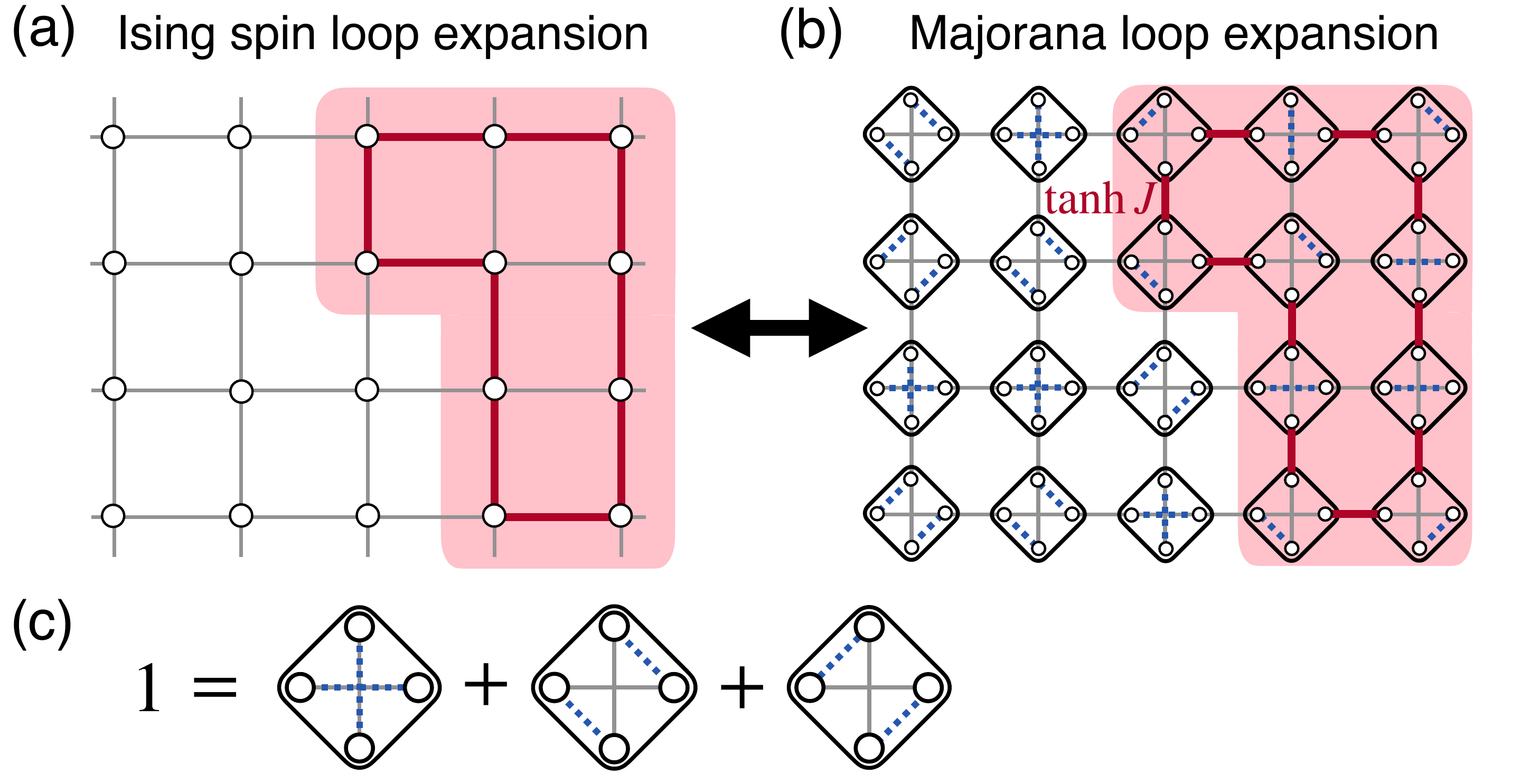}\caption{Representation of the loop gas ensemble via the Ising model or Majorana
fermions. (a) shows a loop from the high-temperature expansion of
the Ising model $Z_{\text{IM}}\left[J\right]\equiv\sum_{\left\{ \sigma\right\} }e^{J\sum_{\langle i,\ j\rangle}\sigma_{i}\sigma_{j}}$,
where the red solid line represents $\tanh\left(J\right)\sigma_{i}\sigma_{j}$.
The closed loop weight is $\left(\tanh J\right)^{\left|\mathcal{C}\right|}$,
with $\left|\mathcal{C}\right|$ denoting the loop length. (b) depicts
the corresponding loop in the Majorana representation, where the red
solid and blue dashed lines represent Majorana bilinears. The loop
has an amplitude of $\left(\tanh J\right)^{\left|\mathcal{C}\right|}$
after integrating out the Majorana spinors, as the onsite Majorana
term is normalized to one for different configurations (c).
\label{fig:fermionization_ising_majorana}}
\end{figure}

In this section, we first derive the Majorana representation for the Ising model following Ref. \cite{samuel1980jmp, wegner2016springer}, and then apply it to R\'enyi-$2$ coherent information (CI).

This fermionization method represents a loop gas ensemble using either the Ising model
or Majorana fermions (see Fig.~\ref{fig:fermionization_ising_majorana}).
The loop gas ensemble consists of closed loops $\mathcal{C}$ of length $\left|\mathcal{C}\right|$,
with partition function, 
\begin{equation}
Z\left[J\right]=\mathcal{N}^\prime\sum_{\left\{ \mathcal{C}\right\} }\left(\tanh J\right)^{\left|\mathcal{C}\right|},\label{eq:loop_gas}
\end{equation}
where $\mathcal{N}^\prime$ is an irrelevant normalization factor, and $\tanh J$
represents the string tension, assigning weight based on loop length
$\left|\mathcal{C}\right|$. This describes the Ising model with partition
function $Z_{\text{IM}}\left[J\right]=\sum_{\left\{ \sigma\right\} }e^{J\sum_{\langle i,\ j\rangle}\sigma_{i}\sigma_{j}}$,
as illustrated in Fig.~\ref{fig:fermionization_ising_majorana}
(a), 

\begin{eqnarray}
Z_{\text{IM}}\left[J\right] & = & \left(\cosh J\right)^{2N}\sum_{\{\sigma\}}\prod_{\langle i,\ j\rangle}\left(1+\tanh J\sigma_{i}\sigma_{j}\right)\nonumber \\
 & = & \left(\cosh J\right)^{2N}\sum_{\left\{ \mathcal{C}\right\} }\left(\tanh J\right)^{\left|\mathcal{C}\right|},
\end{eqnarray}
where only closed loops contribute after summing over the Ising spins.
Alternatively, the loop gas can be represented using Majorana fermions
by attaching four Majorana spinors to each site (see Fig.~\ref{fig:fermionization_ising_majorana}
(b)), connecting nearest neighbor. The partition function is then
\begin{equation}
Z_{\text{MH}}\left[J\right]=\int\mathcal{D}\gamma e^{-\beta\sum_{\boldsymbol{r}_{1},\ \boldsymbol{r}_{2}}\Psi_{\boldsymbol{r}_{1}}^{T}H_{\text{MH},\ \boldsymbol{r}_{1}\boldsymbol{r}_{2}}\Psi_{\boldsymbol{r}_{2}}},\label{eq:path_integral}
\end{equation}
where $\mathcal{D}\gamma=\prod_{\boldsymbol{r}}\prod_{a}d\gamma^{\left(a\right)}$
is the integration measure, $\beta\Psi^{T}H_{\text{MH}}\Psi$ is a Majorana bilinear,
and $\Psi_{\boldsymbol{r}}=\left(\gamma_{\boldsymbol{r}}^{\left(u\right)},\ \gamma_{\boldsymbol{r}}^{\left(d\right)},\ \gamma_{\boldsymbol{r}}^{\left(r\right)},\ \gamma_{\boldsymbol{r}}^{\left(l\right)}\right)^{T}$
is a four-component spinor. Here, $\beta\ll1$ serves as a controlling
parameter, tracking the expansion power. To match Eq.~\eqref{eq:loop_gas},
the Hamiltonian $H_{\text{MH}}$ in Fig.~\ref{fig:Hamiltonian_2d_pwave} (with
$t\equiv\tanh J$) is chosen so that a loop configuration has an amplitude $t^{\left|\mathcal{C}\right|}$ coinciding with the Ising model: The Majorana hopping constant accounts for string tension, and the on-site Majorana terms normalize to one (visualized in Fig.~\ref{fig:fermionization_ising_majorana}
(c)), i.e., 
\begin{equation}
\frac{1}{\beta^2}\int \mathcal{D}\gamma_{\boldsymbol{r}}\ e^{-\beta \Psi^{T}_{\boldsymbol{r}} H_{\text{MH}, \boldsymbol{r}\boldsymbol{r}}\Psi_{\boldsymbol{r}}} = 1.
\end{equation}
Thus, we confirm that $Z_{\text{MH}}$
also represents a loop gas ensemble by expanding in $\beta$ and tracing out $\gamma$ (see Fig. (\ref{fig:fermionization_ising_majorana})
(b) for a exemplary loop configuration).

The partition function $Z_{\text{MH}}[J]$ can be also be recast in the operator formalism, to wit 
\begin{equation}
Z_{\text{MH}}\left[J\right]=\lim_{\beta\rightarrow0^{+}}\text{Tr}\left[e^{-\beta\hat{H}_{\text{MH}}}\left(-1\right)^{\hat{Q}}\right]\equiv \lim_{\beta\rightarrow 0^+}\mathcal{P},
\end{equation}
where the $\beta\rightarrow 0^+$ reflects the classical nature of the underlying spin model. Note that the appearance of the fermion parity operator $\left(-1\right)^{\hat{Q}}$ in the representation of the Majorana partition function as a trace over Hilbert space. It crucially modifies the temporal boundary condition of the Majorana fermions from anti-periodic to periodic, aligning them with the underlying spin representation. This modification can be interpreted as the insertion of a temporal $\mathbb{Z}_2$ flux, a common feature in fermionization/bosonization.

On a torus, the exact mapping between the Ising model and Majorana
fermions is 
\begin{eqnarray}\label{supp_eq:Ising_torus}
 &  & 2Z_{\text{IM}}\left[J\right]\nonumber \\
 & = & \lim_{\beta\rightarrow0^{+}}\frac{1}{\beta^{2N}}\left(-\mathcal{P}_{\text{PP}}\left[\beta,\ \tanh J;\ \eta\right]+\mathcal{P}_{\text{AA}}\left[\beta,\ \tanh J;\ \eta\right]\right)\nonumber \\
 &  & +\lim_{\beta\rightarrow0^{+}}\frac{1}{\beta^{2N}}\left(\mathcal{P}_{\text{PA}}\left[\beta,\ \tanh J;\ \eta\right]+\mathcal{P}_{\text{AP}}\left[\beta,\ \tanh J;\ \eta\right]\right),\nonumber \\
\end{eqnarray}
and 
\begin{equation}\label{supp_eq:Ising_torus_2}
\sum_{\alpha}Z_{\text{IM},\ \alpha}\left[J\right]=\lim_{\beta\rightarrow0^{+}}\frac{1}{\beta^{2N}}\sum_{\alpha}\mathcal{P}_{\alpha}\left[\beta,\ \tanh J;\ \eta\right],
\end{equation}
with $\eta=+1$.
The derivation is intricate due to non-contractible loops on the torus, which complicate the Majorana representation for the loop gas ensemble  and require careful handling of different boundary-condition sectors. We refer interested readers to Ref. \cite{wegner2016springer} for further details. Notably, the different spatial boundary condition sectors (labeled by $\alpha$) on the right-hand side of Eq.~(\ref{supp_eq:Ising_torus}, \ref{supp_eq:Ising_torus_2}), correspond to the insertion of a $\mathbb{Z}_2$ spatial flux, which combined with the fermion parity operator, demonstrates that the Ising model maps to a Majorana model with both temporal and spatial $\mathbb{Z}_2$ fluxes.

Finally, as an illustration, it is instructive to apply this mapping
to the R\'enyi-$2$ CI, which is mapped to a clean and
non-interacting Majorana model, i.e., 
\begin{equation}
I_{c}^{\left(2\right)}=2\log_{2}\left(1-2\frac{\mathcal{P}_{\text{PP}}\left[\beta,\ t_{2};\ \eta\right]}{\sum_{\alpha}\mathcal{P}_{\alpha}\left[\beta,\ t_{2};\ \eta\right]}\right),\ t_{2}=\left(1-2p\right)^{2},
\end{equation}
with $\eta=+1$.
At finite system sizes, the gap closing occurs only in the `PP'
sector,
at $t_{2}=\sqrt{2}-1$  (see Fig. \ref{fig:Hamiltonian_2d_pwave} for numerical results), consistent with the results from self-duality.

\begin{figure}[t!]
\includegraphics[scale=0.18]{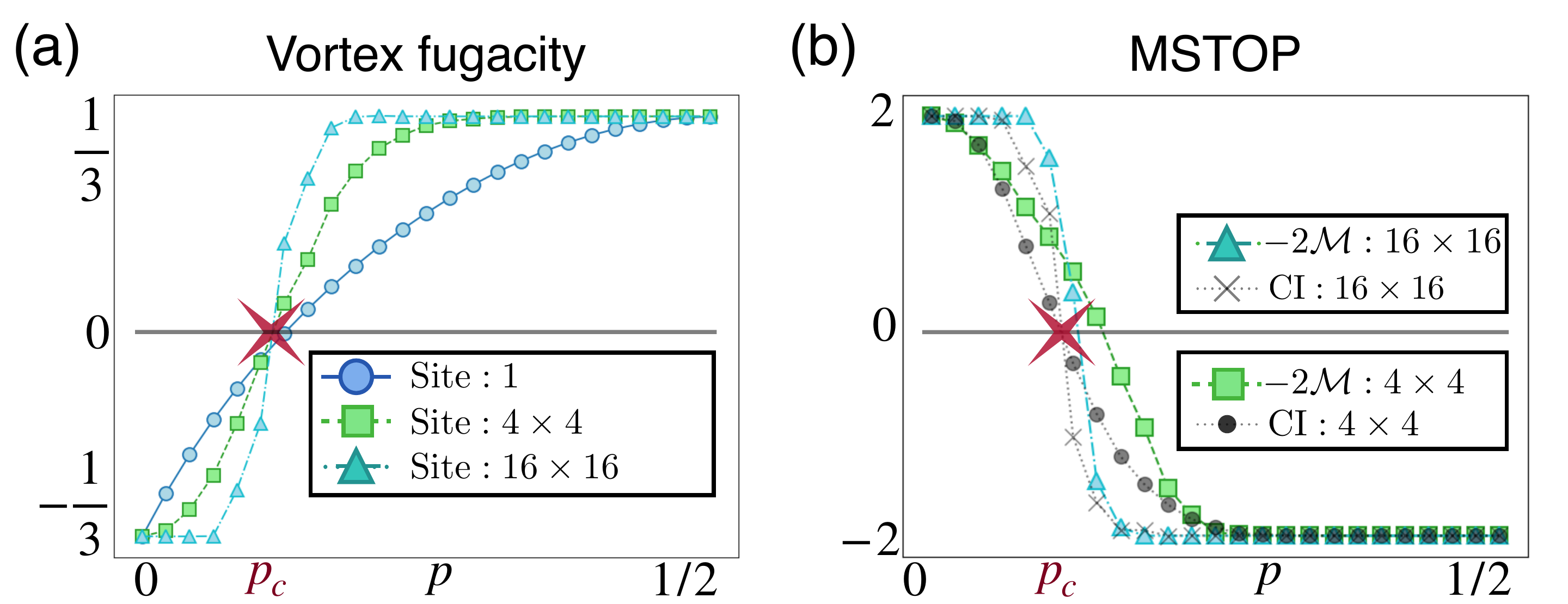}\caption{Numerical results for vortex fugacity $u$ (a) and mixed-state topological order parameter $\mathcal{M}$ (b). (a) shows $u$ as negative/positive in the topological/normal phase ($p \leq p_c$/$p > p_c$), vanishing at the critical point with small finite size dependence, making it a useful probe for identifying the critical point in small systems. (b) shows the disorder average mixed-state topological order parameter (MSTOP), $-2\mathcal{M}$, across different sizes, along with coherent information (CI), confirming that $-2\mathcal{M}$ aligns with CI in the gapped phase but exhibits size dependence near the critical point, as $\mathcal{M}$ is defined for the gapped phase (Ref. \cite{huang2024arxiv}).
\label{supp_fig:fugacity_msto}}
\end{figure}

\section{Additional numerical results on topological phase transitions in class D Majorana representation \label{supp_sec:numerics_tpt}}
We provide additional numerical results for the decoding phase transition, from the perspective of a class D Majorana (see Fig. \ref{supp_fig:fugacity_msto}), namely the vortex fugacity, and the mixed-state topological order parameter.
\begin{itemize}
\item Vortex fugacity, defined as the ratio
\begin{equation}
u \equiv \langle \langle \frac{\mathcal{P}_{\text{PP}}[\beta,\ t_2;\ \eta]}{\mathcal{P}_{\text{AP}}[\beta,\ t_2;\ \eta]+\mathcal{P}_{\text{PA}}[\beta,\ t_2;\ \eta]+\mathcal{P}_{\text{AA}}[\beta,\ t_2;\ \eta]}\rangle \rangle,
\end{equation}
with the limit $\beta \rightarrow 0^+$ taken hereafter. Around the critical point, the CI relates to $u$ by expanding $\mathcal{P}_{\text{PP}}[\beta, t_2; \eta]$ and to leading order,
\begin{equation}
I_c=-\frac{4}{\ln 2}\left(\lim_{\beta\rightarrow 0^+}u\right) +\ \mathcal{O}\left[(\mathcal{\mathcal{P}_{\text{PP}}})^2\right].
\end{equation}
Numerical results in Fig. \ref{supp_fig:fugacity_msto} (a) confirm that $u$ has small finite size effects at the zero-fugacity point.

\item Disorder averaged mixed-state topological order parameter (MSTOP) 
\begin{equation}
\mathcal{M}\equiv\langle\langle\text{sign}\left(\frac{\mathcal{P}_{\text{PP}}\left[\beta,\ t_{1};\ \eta \right]\mathcal{P}_{\text{AA}}\left[\beta,\ t_{1};\ \eta\right]}{\mathcal{P}_{\text{AP}}\left[\beta,\ t_{1};\ \eta\right]\mathcal{P}_{\text{PA}}\left[\beta,\ t_{1};\ \eta\right]}\right)\rangle\rangle,
\end{equation}
with its sign representing a mixed-state topological order parameter \cite{huang2024arxiv}.
Numerical results in Fig. \ref{supp_fig:fugacity_msto} (b) show that $\mathcal{M}$ matches with the CI in the gapped phases, where this quantity is well defined. 

\end{itemize}
\bibliographystyle{apsrev4-2}

\begin{thebibliography}{75}
\bibitem [{fnC()}]{fnCI}%
  \BibitemOpen
  \href@noop {} {}\bibinfo {note} {In practice, we determine the error
  threshold from the coherent information crossing point across different
  system sizes. This is because depending on the code model, the crossing point
  may deviate from zero due to noise-induced offsets; for example, in the toric
  or color code with bit-flip errors, it is located at $1/2$.}\BibitemShut
  {Stop}%
\bibitem [{fnG()}]{fnGibbsToric}%
  \BibitemOpen
  \href@noop {} {}\bibinfo {note} {This contrasts with the toric code at finite
  temperature, where all error chains have the same energy cost regardless of
  length.}\BibitemShut {Stop}%
\end{thebibliography}

\end{document}